\documentclass[aps,floats,showpacs,nofootinbib,floatfix]{revtex4}

\usepackage{amsfonts}
\usepackage{amsmath}
\usepackage{graphicx}
\usepackage{slashed}

{\catcode`\|=\active \gdef\Braket#1{\left<\mathcode`\|"8000\let|\bravert
  {#1}\right>}}
\def\bravert{\egroup\,\vrule\,\bgroup}
\newcommand{\alltrace}{\ensuremath{\mathbb{T}\mathrm{r}}}

\begin{document}

\title{Generalized parton distributions of the pion in a Bethe-Salpeter
approach}
\author{S. Noguera}
\email{Santiago.Noguera@uv.es}
\author{L. Theu{\ss }l}
\email{theussl@triumf.ca}
\altaffiliation[Present address: ]{TRIUMF, Vancouver, B.C. V6T 2A3, Canada }
\author{V. Vento}
\email{Vicente.Vento@uv.es}
\affiliation{Departamento de F\'{\i}sica Te\`orica and
Instituto de F\'{\i}sica Corpuscular,
Universidad de Valencia - CSIC, E-46100 Burjassot (Valencia), Spain.}
\date{\today }

\begin{abstract}
We calculate generalized parton distribution functions in a field
theoretic formalism using a covariant Bethe-Salpeter approach for
the determination of the bound-state wave function. We describe
the procedure in an exact calculation in scalar Electrodynamics
proving that the relevant corrections outside our scheme vanish.
We extend the formalism to the Nambu--Jona-Lasinio model, a
realistic theory of the pion. We go in both cases beyond all
previous calculations and discover that all important features
required by general physical considerations, like symmetry
properties, sum rules and the polynomiality condition, are
explicitly verified. We perform a numerical study of their
behavior in the weak and strong coupling limits.
\end{abstract}

\maketitle



\section{Introduction}

\noindent Hard reactions provide important information for unveiling the
structure of hadrons. The large virtuality, $Q^{2}$, involved in these
processes allows the factorization of the hard (perturbative) and soft
(non-perturbative) contributions in their amplitudes. Therefore these
reactions are receiving great attention by the hadronic physics community.
Among the hard processes, the Deeply Virtual Compton Scattering (DVCS)
merits to be singled out, because it can be expressed, in the asymptotic
regime, in terms of the so called generalized parton distributions
(GPDs)~\cite{Muller:1994fv, Ji:1997nm,Radyushkin:1997ki,Ji:1998pc}. The GPDs
describe non-forward matrix elements of light-cone operators and therefore
measure the response of the internal structure of the hadrons to the
probes~\cite{Diehl:1998kh,Diehl:2000xz, Polyakov:1999gs,Diehl:1999ek}.
There is much effort under way related to the measurement of these functions.

Due to the impossibility at present to determine the GPDs from Quantum
Chromodynamics directly, models have been used to provide estimates which
should serve to guide future
experiments~\cite{Ji:1997gm,Anikin:2001zv,Petrov:1998kf,Penttinen:1999th,Scopetta:2002mz,Scopetta:2002xq}.
The aim of our work is to perform such a
calculation in a field theoretic scheme which treats the bound-state in a
fully covariant manner following the Bethe-Salpeter approach. In this way we
would like to preserve all invariances of the problem. For simplicity we
shall use mesons as initial and final states.

We define a scheme, to calculate the electro-magnetic interaction of
hadrons, which separates the soft parts, where we use a non perturbative
treatment, and the hard parts, where the conventional perturbative treatment
is applied. The scheme preserves gauge invariance to leading order and
leading twist.

In order to describe the soft part we start by using as a test of
our ideas a model based on the $\phi ^{4}$ field theory. This
theory has the advantage of being renormalizable and therefore any
contribution can be analyzed properly. We show, for example, that
a correction to the hand bag contribution to first order in the
strong coupling constant, which goes beyond our scheme, vanishes.

We next proceed to the main development of this paper, namely to
perform the study of the GPDs of the pion by using the
Nambu--Jona-Lasinio (NJL) model to describe its structure. The NJL
model is not a toy model. In fact, it is the most realistic model
for the pion based on a quantum field theory built with quarks. It
gives a good description of the low energy physics of the
pion and respects the realization of chiral symmetry~\cite{Klevansky:1992qe}.
Moreover it has been used as the model to tune many coefficients of Chiral
Perturbation Theory~\cite{Bijnens:1996ww}.

The NJL model is a non renormalizable field theory and therefore a
cut-off procedure has to be defined. We have chosen the
Pauli-Villars regularization procedure because it respects all the
symmetries of the problem. In our scheme, we use the NJL model to
describe the soft (non perturbative) part of the process, i.e. the
initial and final states, while for the hard part we use
conventional perturbative QCD. The use of the NJL model allows to
calculate the GPDs for massive pions. Some peculiarities, as the
non vanishing of the GPDs at the kinematic boundary regions, $x=0$
and $x=1$, well known in the exact $m_\pi = 0$ limit, survive when
$m_\pi \ne 0$.

Our paper is organized as follows. In section II we give some general
definitions for scalar partons and we introduce our kinematical variables.
In section III and IV we will define our approach for the scalar and the NJL
model respectively. Section V presents our results and Section VI our
conclusions.


\section{Generalized parton distributions}

\begin{figure}[tbp]
\begin{center}
\includegraphics[width=20em]{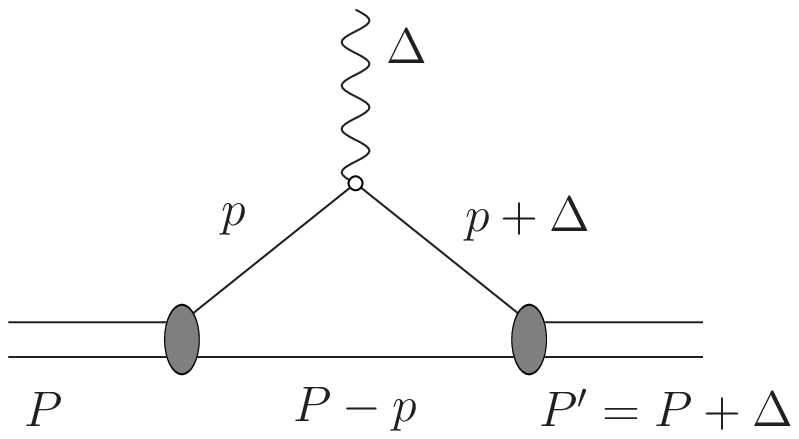} \hspace{2em} %
\includegraphics[width=20em]{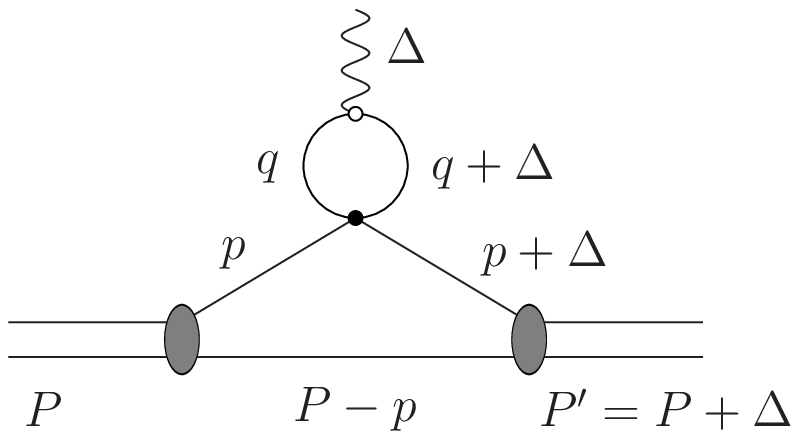}
\end{center}
\caption{Diagrams up to order $g$ for the GPDs. The black dot
indicates the scalar interaction vertex, while the white blob
indicates the effective quark -- gauge-boson vertex.} \label{fig1}
\end{figure}

\noindent The GPDs are non-diagonal matrix elements of bi-local field
operators. Various conventions, reference frames, variables, etc., have been
used in the literature for the description of such objects. Our notation is
represented in Fig.~\ref{fig1}, i.e., the initial momentum is labeled by $P$%
, the final momentum by $P^{\prime }$, and the momentum transfer is given by
$\Delta =P^{\prime }-P$. We shall describe initially a model with scalar
particles, the generalization to particles with spin is straightforward and
will be outlined in a later section. The GPD of a scalar system is defined
by the matrix elements of bi-local scalar field
operators~\cite{Muller:1994fv,Ji:1997nm,Radyushkin:1997ki,Ji:1998pc}:
\begin{equation}
\mathcal{J}^{+}\equiv \frac{1}{2}\left. \int \frac{dz^{-}}{2\pi }%
e^{ixP^{+}z^{-}}\left\langle P^{\prime }\right\vert \Phi ^{\dagger }\left(
0\right) \overset{\leftrightarrow }{\partial}{}^{+}\Phi \left( z\right)
\left\vert P\right\rangle \right\vert _{z^{+}=z^{\perp }=0}=\mathcal{H}%
(x,\zeta ,t),  \label{spdgen}
\end{equation}
where $x$ is the conventional Bjorken variable, $\zeta $ the so-called
skewedness parameter, and $\overset{\leftrightarrow }{\partial }=\overset{%
\rightarrow }{\partial }-\overset{\leftarrow }{\partial }$. The elastic
electromagnetic form factor of a system composed of two scalar particles is
given by:
\begin{equation}
J^{+}\equiv \left\langle P^{\prime }\right\vert \Phi ^{\dagger }\left(
0\right) \overset{\leftrightarrow }{\partial}{}^{+}\Phi \left( 0\right)
\left\vert P\right\rangle =(P+P^{\prime })^{+}\,F(t).  \label{emffgen}
\end{equation}
It follows directly from these definitions that integrating the GPD over $x$
gives the form factor,
\begin{equation}
\frac{2}{2-\zeta }\int \mathcal{H}(x,\zeta ,t)\,dx=F(t),  \label{relat}
\end{equation}
where the dependence on the skewedness parameter $\zeta $ drops
out. This result is an important constraint for any model
calculation.

Let us now present our notation. Any four vector $v^\mu$ will be denoted $%
(v^+,v^{\perp},v^-)$, where the light cone variables are defined by $v^\pm =
(v^0\pm v^3)/\sqrt{2}$ and the transverse part $v^{\perp}=(v^1,v^2)$. For
the kinematics indicated in Fig.~\ref{fig1}, we introduce the ratios
\begin{equation}  \label{ratios}
x=\frac{p^+}{P^+}\, , \qquad \qquad \qquad \zeta=-\frac{\Delta^+}{P^+}
\end{equation}
of plus-components. With these definitions, which differ from other
conventional ones~\cite{Golec-Biernat:1998ja,Guichon:1998xv}, both $x$ and $%
\zeta$ are defined on the interval $[0,1]$.

We are only going to consider elastic processes, so $P^{2}=P^{\prime
2}=M^{2} $ and $\Delta ^{2}=t$. The following relation is true in general:
\begin{equation}
\left( \Delta ^{\perp }+\zeta P^{\perp }\right) ^{2}=-\zeta
^{2}M^{2}-(1-\zeta )t,  \label{tdef}
\end{equation}%
the positivity of which implies an upper bound for the skewedness $\zeta $
at a given value of the momentum transfer $t$: $\zeta \leq
(-t)/(2M^{2})\left( \sqrt{1+4M^{2}/(-t)}-1\right) \leq 1$.


\section{Scalar Electrodynamics}

\noindent We begin our study with the simplest model which allows for a
completely analytic solution of the Bethe-Salpeter equation: this model
describes a bound state of two distinguishable equal-mass scalar particles
bound together by a zero-range interaction. For later convenience, we choose
only one of the constituent particles to be charged. Our Lagrangian is,
\begin{equation}
\mathcal{L}=\left[ D_{\mu }\phi \right] ^{\dagger }\left[ D^{\mu }\phi %
\right] -m^{2}\phi ^{\dagger }\phi +\frac{1}{2}\partial _{\mu }\chi \partial
^{\mu }\chi -\frac{1}{2}m^{2}\chi ^{2}-\frac{g}{2}\left( \phi ^{\dagger
}\phi \,\chi ^{2}\right) ,  \label{ZRlag}
\end{equation}
with $D_{\mu }=\partial _{\mu }+ieA_{\mu }$ so that the electromagnetic
charge only couples to the field $\phi $. Assuming that the coupling
constant $g$ is larger than a critical value, we have bound states stemming
from the last term in Eq.~(\ref{ZRlag}). The corresponding Bethe-Salpeter
equation is trivially solved in the ladder approximation~\cite{Itzykson:1985}%
, providing the fully covariant amplitude for the bound state of total mass $%
M^{2}=P^{2}$:
\begin{equation}
\Phi (P,p)=\frac{C}{\left( p^{2}-m^{2}+i\epsilon \right) \left( \big(P-p\big)%
^{2}-m^{2}+i\epsilon \right) },  \label{zram}
\end{equation}
where $p$ is the four-momentum of one of the two constituents with equal
mass $m$, and $C$ is a normalization constant. The spectrum for the ground
state in this model, can be obtained directly from the Bethe-Salpeter
equation, and is given by
\begin{equation}
1=-igI_{2}(P),  \label{specscal}
\end{equation}%
where the integral $I_{2}(P)$ is given by Eq.~(\ref{int2}). Note that this
integral is actually divergent, so some sort of renormalization would be
required in order to calculate the spectrum. The theory defined by
Eq.~(\ref{ZRlag}) is renormalizable and a renormalization program for bound states
can be defined. However, for the evaluation of parton distribution functions
in this model, we do not encounter any divergent integrals, we shall
therefore ignore any matters of renormalization.

The model defined by the above equations is certainly not realistic enough
to furnish a reasonable description of real, physical bound states, like the
pion for instance. Its main advantage lies in its simplicity, the fact that
one may obtain analytic solutions, avoiding approximations that might
destroy physical requirements or symmetries, like Lorentz- or Gauge
invariance, sum rules, etc. Moreover it has the added advantage that it is
defined within a renormalizable quantum field theory. These properties make
it a useful playground to perform benchmark calculations, as it was used
recently in order to test the viability of certain relativistic quantum
mechanics approaches~\cite{Desplanques:2001ze,Amghar:2002jx}.

\subsection{Generalized parton distribution function}

\noindent Our starting point is to write the generalized parton distribution
(GPD) as an integral over Bethe-Salpeter amplitudes. To do so we could use
Eq.~(\ref{spdgen}), a procedure which we shall develop in the NJL case.
However for completeness we follow here the procedure of
ref.\cite{Tiburzi:2002tq,Tiburzi:2001ta}, which establishes a relation between the
calculation of the GPD and that of the electromagnetic form factor. Namely,
in the scattering process $\gamma^* + \pi \rightarrow \gamma + \pi$, the
leading ``hand-bag'' diagram, in the deep inelastic limit, reduces to a
triangle diagram with an effective vertex (containing two-photon
contributions), resulting from the contraction of the propagator with
infinite momentum (see first diagram in Fig.~\ref{fig1}). Then,
\begin{eqnarray}
\mathcal{H}(x,\zeta ,t) &=&-\frac{i}{2}\int \frac{d^{4}p}{(2\pi )^{4}}\delta
(p^{+}-xP^{+})\,\bar{\Phi}(P,p)(\Delta +2p)^{+}\left[ (P-p)^{2}-m^{2}\right]
\Phi (P+\Delta ,p+\Delta )  \notag  \label{spdfbs} \\
&=&\frac{2x-\zeta }{2}\,\frac{1}{i}\int \frac{d^{4}p}{(2\pi )^{4}}\delta
(x-p^{+}/P^{+})\,\bar{\Phi}(P,p)\left[ (P-p)^{2}-m^{2}\right] \Phi (P+\Delta
,p+\Delta ).
\end{eqnarray}
Note that the adjoint Bethe-Salpeter amplitude $\bar{\Phi}$ is defined via
an anti-chronological time ordering of field operators~\cite{Itzykson:1985}
as compared to $\Phi $, which means that the infinitesimal part $i\epsilon$
in Eq.~(\ref{zram}) keeps the same sign as compared to the mass term $m^{2}$.

Using the Bethe-Salpeter amplitude of Eq.~(\ref{zram}), we obtain
\begin{equation}
\mathcal{H}(x,\zeta ,t)=\frac{2x-\zeta }{2}\,\frac{C^{2}}{i}\int \frac{d^{4}p%
}{(2\pi )^{4}}\frac{\delta (x-p^{+}/P^{+})}{(p^{2}-m^{2}+i\epsilon )\left[
(p+\Delta )^{2}-m^{2}+i\epsilon \right] \left[ (P-p)^{2}-m^{2}+i\epsilon %
\right] }.  \label{spdfbsp}
\end{equation}
Inspecting the pole structure of the integrand for the evaluation of the $%
p^{-}$ integral, we note that it vanishes unless $0\leq x\leq 1$, i.e., the
GPDs have the correct support properties. The integral of Eq.~(\ref{spdfbsp}%
) may be calculated explicitly. The analytic result for the GPD is then
simply
\begin{equation}
\mathcal{H}(x,\zeta ,t)=\frac{C^{2}}{16\pi ^{2}}\frac{2x-\zeta }{2}\,\tilde{I%
}_{3}(m,x,\zeta ,t),\qquad 0\leq x\leq 1,  \label{anares}
\end{equation}
with $\tilde{I}_{3}(m,x,\zeta ,t)$ given by Eq.~(\ref{regul4}). We observe
that our result is explicitly covariant, depending only on $x$, $\zeta $ and
$t$. Note that the complete solution is explicitly continuous at $x=\zeta $,
however, the derivative at this point is discontinuous. We furthermore
encounter a zero at the point $x=\zeta /2$, which is due to the photon
vertex, see Eq.~(\ref{spdfbs}). The quark distribution function is given by
\begin{equation}
q(x)\equiv \mathcal{H}(x,0,0)=\frac{C^{2}}{16\pi ^{2}}\frac{x(1-x)}{%
m^{2}-x(1-x)M^{2}}.  \label{zrqdf}
\end{equation}
The normalization integral may be done analytically and determines the
normalization constant $C$.

As it is not a common practice to write GPDs as integrals over
Bethe-Salpeter amplitudes, we note that we can write $\mathcal{H}(x,\zeta
,t) $ for $x>\zeta $ as the product of light cone wave functions, defined by:
\begin{equation}
\Psi (x,p^{\perp })\equiv \frac{P^{+}\sqrt{x(1-x)}}{i\pi }\int dp^{-}\,\Phi
(P,p)=-\frac{C\sqrt{x(1-x)}}{{\left( p^{\perp }-xP^{\perp }\right) }%
^{2}+m^{2}-x(1-x)M^{2}},  \label{bslf}
\end{equation}
which is non-zero only for $0<x<1$. Using some kinematic relations, the GPD
for $x>\zeta $ may then be written as
\begin{equation}
\left. \mathcal{H}(x,\zeta ,t)\rule[-1ex]{0em}{2ex}\right\vert _{x>\zeta }=%
\frac{1}{2}\,\frac{2x-\zeta }{2}\,\frac{1}{\sqrt{x(x-\zeta )}}\int \frac{%
d^{\,2}p^{\perp }}{(2\pi )^{3}}\Psi ^{\ast }\left( \frac{x-\zeta }{1-\zeta }%
,p^{\perp }+\frac{1-x}{1-\zeta }\Delta ^{\perp }\right) \Psi \left(
x,p^{\perp }\right) .  \label{xgtxi}
\end{equation}
On the other hand, for $x<\zeta $, the GPD may not directly be written as
the product of light front wave functions like in Eq.~(\ref{xgtxi}), even
though we are still able to find analytic solutions for the integral of
Eq.~(\ref{spdfbsp}).

Another conventional way of writing GPDs is via a parameterization in form
of double distributions~\cite{Polyakov:1999gs}. The relation of these double
distributions and the associated D-term with the results of a scalar model
similar to the one considered here was recently discussed in
ref.~\cite{Tiburzi:2002tq}.

\begin{figure}[tbp]
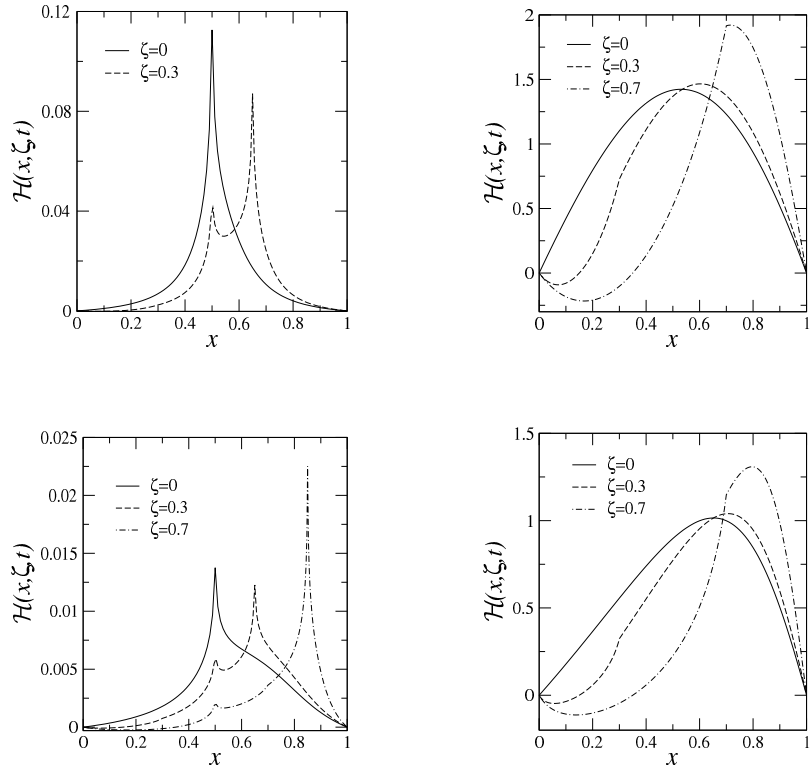

\begin{center}
\includegraphics*[width=14em]{Paper_Def_SCL_wea_01.eps}
\hspace{1.5cm}
\includegraphics*[width=13.5em]{Paper_Def_SCL_140_01.eps}
\\
\vspace{1.0cm}
\includegraphics*[width=14em]{Paper_Def_SCL_wea_10.eps}
\hspace{1.5cm}
\includegraphics*[width=13.5em]{Paper_Def_SCL_140_10.eps}
\end{center}
\caption{Generalized parton distributions in the scalar model for
different values of the bound state mass $M$ and momentum
transfers $t$. The top row gives results for $t= -0.1$ GeV$^2$.
The lower row those for $t= -1$ GeV$^2$. The graphs follow the
scenarios discussed in the text: i) (weak binding, $M\approx 2m$) left column
and ii) (deep binding, $M = m_\pi$) right column.
The values of the third variable $\zeta$ are shown in the figures.
Note that the sum rule of Eq.~(\protect\ref{relat}) is
exactly satisfied for each graph.} \label{figsed}
\end{figure}

In Fig.~\ref{figsed} we give examples of GPDs in this model for
different values of the binding energy and momentum transfers. We
distinguish two scenarios: i) the weak binding scenario $M\approx
2m$ (up to small binding energies of electromagnetic magnitude);
ii) the deep binding scenario, which we choose close to the chiral
limit, $M = m_\pi \approx 140$ MeV, for reasons which will become
natural in our later study of the Nambu--Jona-Lasinio model.

We checked numerically that the sum rule of Eq.~(\ref{relat}) is
always exactly satisfied. A more stringent test is the so-called
polynomiality condition~\cite{Ji:1998pc}, which states that the
moments of the GPDs,
\begin{equation}
\int_{0}^{1}dx\;x^{N-1}\mathcal{H}(x,\zeta ,t)\equiv F_{N}(\zeta ,t),
\label{poly}
\end{equation}
give functions $F_{N}(\zeta ,t)$ that are polynomials in $\zeta $ of order
not higher than $N$. It is difficult to verify this analytically in the
general case, even with the exact solutions that we obtained. Just in the
limiting case $M^{2}=0$, the following relation may be shown to hold:
\begin{equation}
F_{N}(\zeta ,0)=3\frac{2N+(N-2)\sum_{i=1}^{N}\zeta ^{i}}{N(N+1)(N+2)},
\label{zrpoly}
\end{equation}
i.e., the polynomiality condition is exactly satisfied at zero momentum
transfer. Note also that for $N=1$, Eq.~(\ref{zrpoly}) gives the correct
normalization of Eq.~(\ref{relat}) for any value of $\zeta $. This shows
explicitly that the sum rule is indeed independent of $\zeta $ in this case.
For $t\neq 0$ and $M^{2}\neq 0$ we have shown, by numerical integration,
that the polynomiality condition also holds (see appendix~\ref{polynomiality}).

In the small binding limit, $M^{2}\rightarrow 4m^{2}$, we recover
the peaked nature of the GPDs, that was observed in
ref.~\cite{Tiburzi:2001ta}, where also an interpretation was
given. For small binding energies, the constituents are almost
free, so their wave function is highly peaked in momentum space.
The two peaks at $x=1/2$ and $x=(1+\zeta )/2$ correspond just to
the maxima of the corresponding wave function in the overlap
formula Eq.~(\ref{xgtxi}). In the exact limit $M = 2m$, the GPDs
reduce to a sum of two $\delta $-functions\footnote{ Note that the
form factor in this limit is equal to zero everywhere except at
$t=0$, where it is 1.}:
\begin{equation}
\mathcal{H}(x,\zeta ,t)=\frac{1}{2}\left[ \delta \left( x-\frac{1}{2}\right)
+\delta \left( x-\frac{\zeta +1}{2}\right) \right] \frac{2-\zeta }{2}F(t).
\label{deltaGPD}
\end{equation}
In particular, the quark distribution function $q(x)=\mathcal{H}(x,0,0)$
becomes simply $q(x)=\delta (x-1/2)$, which means that the particles are
free.

In the deep binding limit, $M = m_\pi$, the GPDs change from the
sharp peaks of the previous case to broad bumps describing the
Fermi motion of the deeply bound constituents. At zero momentum
transfer, $t=0$, we find the simple form
\begin{equation}
\mathcal{H}(x,\zeta ,0)=6\,\frac{2x-\zeta }{2}\left\{ \frac{x}{\zeta }\theta
\left( \zeta -x\right) \,\theta \left( x\right) +\frac{1-x}{1-\zeta }\theta
\left( x-\zeta \right) \,\theta \left( 1-x\right) \right\} .  \label{M0t0sca}
\end{equation}

\subsection{Electromagnetic form factor}

\noindent
Calculating the electromagnetic form factor according to Eq.~(\ref{emffgen}),
we find:
\begin{equation}  \label{zrff1}
(P+P^{\prime})^\mu\,F(Q^2)=iC^2 \int \frac{d^{4}p}{(2\pi)^4} \frac{%
(\Delta+2p)^\mu}{(p^2-m^2 + i\epsilon)\left[ (p+\Delta)^2-m^2 + i\epsilon%
\right] \left[ (P-p)^2-m^2 + i\epsilon\right]},
\end{equation}
leading to~\cite{Amghar:2002jx},
\begin{equation}
F(Q^{2})=\frac{C^{2}}{16\pi ^{2}}\int_{0}^{1}dz\,\frac{2}{Q}\frac{1-z}{\sqrt{%
D}}\log \frac{\sqrt{D}+zQ}{\sqrt{D}-zQ}\equiv \frac{2}{2-\zeta }%
\int_{0}^{1}dx\,\mathcal{H}(x,\zeta ,t),  \label{zrff3}
\end{equation}
with $D=4[m^{2}-z(1-z)M^{2}]+z^{2}Q^{2}$. This suggests that one could
calculate the GPD by a simple change of the integration variable in the
expression for the form factor. The general character of this change of
variable would not be clear, however. Naturally, at $Q^{2}=0$ we reproduce
the normalization condition.

Taking the parameters $m = 241\,\mathrm{MeV}$ and $M = 139\,\mathrm{MeV}$,
we can evaluate the root-mean squared radius of a ``pion'' built of scalar
particles, via $\left<r_\pi \right>^2=-6\,
\left.\partial F(Q^2)/\partial Q^2\right|_{Q^2=0} = (0.47\,\mathrm{fm})^2$,
to be compared to the experimental value of
$(0.66 \,\mathrm{fm})^2$. In the limit of a mass-less bound state,
$M = 0$, we obtain
the analytic result $\left<r_\pi\right>^2 = 3/(10\,m^2)=(0.45 \,\mathrm{fm})^2$%
.

\subsection{Bubble diagram and Gauge invariance}

\begin{figure}[tbp]
\begin{center}
\includegraphics[width=9em]{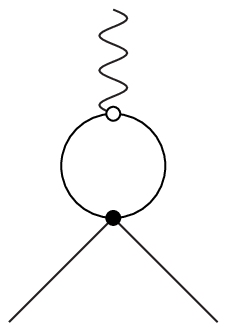} \hspace{1em} \raisebox{9ex}{=}
\hspace{1em} \includegraphics[width=9em]{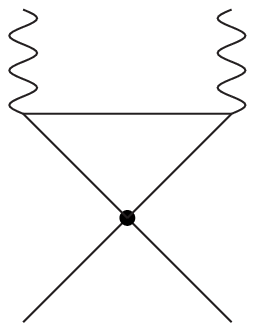} \hspace{1em} %
\raisebox{9ex}{+} \hspace{1em} \includegraphics[width=9em]{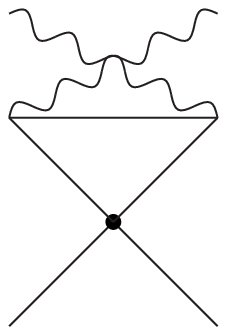}
\hspace{1em} \raisebox{9ex}{+} \hspace{1em} %
\includegraphics[width=9em]{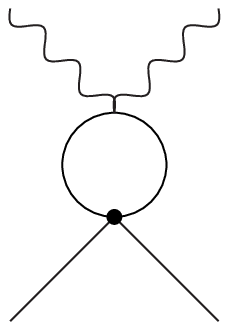} \hspace{1em} \raisebox{9ex}{=}
\hspace{1em} \raisebox{9ex}{0}
\end{center}
\caption{The second diagram of Fig.~\protect\ref{fig1} contains the loop
diagram shown on the left, which corresponds to the expanded set of diagrams
on the right, including the seagull term. All of them are divergent, but
their sum is not. Moreover when we sum them exactly also there finite parts
vanish. The reason behind this cancellation is that the charge is not
renormalized to order~$g$.}
\label{fig2}
\end{figure}

\noindent Up to first order in the strong coupling constant $g$, we have to
calculate the second Feynman diagram of Fig.~\ref{fig1}. In scalar
Electrodynamics, the second diagram does not contribute to the
electromagnetic form factor. The loop integral is proportional to
\begin{equation}  \label{prop}
\int \frac{d^4 q}{(2\pi)^4} \frac{(2q+\Delta)^\mu}{\left(q^2-m^2+i\epsilon%
\right) \left((q+\Delta)^2-m^2+i\epsilon\right)},
\end{equation}
which by virtue of Eq.~(\ref{int6}) is zero. However, this diagram could
contribute to the GPDs, because here we keep the $+$ component of the
integration variable fixed. In a naive way we observe that the bubble
present in the diagram will lead to divergent contributions to the GPDs. In
order to understand properly the problem let us consider the DVCS on the
pion. Unfolding the procedure introduced in this section, we note that the
bubble corresponds to a triangle in which one of the propagators has been
contracted due to the infinite momentum carried by the parton. Divergences
appear in the calculation and correspond in practice to terms proportional
to $\log (Q^2)$ in the triangle diagram. Considering all contributions to
the same order (see Fig.~\ref{fig2}), we find that the divergences cancel.
But not only the divergent terms cancel, also the finite parts do, therefore
the whole contribution, contrary to expectations, vanishes exactly for the
GPDs.

It is not difficult to advance a physical argument explaining this
cancellation. The diagrams of Fig.~\ref{fig2} can be interpreted as a
renormalization of the bare charge present in the seagull term. Due to gauge
symmetry this charge must be the square of the charge associated to the one
photon vertex, before and after renormalization. But we have seen from
Eq.~(\ref{prop}) that there is no renormalization (finite or infinite) at order
$g$ to the charge. Consequently the sum of diagrams of Fig.~\ref{fig2}, or
the second diagram of Fig.~\ref{fig1}, must vanish.


\section{The model of Nambu and Jona-Lasinio}

\noindent The model of the last section can certainly not provide
a very realistic description of electromagnetic properties of a
pion because of the scalar nature of the constituents involved. We
are therefore going to consider a model with spinor particles. An
evident choice to investigate is
the model of Nambu and Jona-Lasinio
(NJL)~\cite{Nambu:1961tp,Nambu:1961fr,Vogl:1991qt,Klevansky:1992qe}. This model is the
most realistic model for pions built of quarks that is based on a quantum
field theory. It gives a good description of the low energy physics of the
pion exhibiting the phenomena of dynamical mass generation and spontaneous
breaking of chiral symmetry, which are crucial ingredients for low-energy
hadronic physics~\cite{Klevansky:1992qe,Bijnens:1996ww}.

We start from the Lagrangian density in the two-flavor version of this model
where we add the electromagnetic interaction in the usual way,
\begin{equation}
\mathcal{L}=\bar{\psi}(iD\hspace{-0.65em}\raisebox{0.2ex}{/}-\mu _{0})\psi +g%
\left[ \left( \bar{\psi}\psi \right) ^{2}-\left( \bar{\psi}\vec{\tau}\gamma
_{5}\psi \right) ^{2}\right] ,  \label{NJLlagem}
\end{equation}
with $D_{\mu }=\partial _{\mu }+ieA_{\mu }$. Again, assuming the contact
interaction of the last term to be mainly responsible for the binding, the
Bethe-Salpeter equation in ladder approximation to be fulfilled by a bound
state in this model is given by (the factor $2$ comes from the symmetry of
the interaction):
\begin{equation}
S^{-1}(p)\,\vec{\Phi}(P,p)\,S^{-1}(p-P)=ig\vec{\tau}\gamma _{5}\int
2\alltrace\left\{ \rule[-1ex]{0em}{3ex}\vec{\tau}\gamma _{5}\cdot \vec{\Phi%
}(P,p^{\prime })\right\} \frac{d^{4}p^{\prime }}{(2\pi )^{4}}.  \label{NJLBS}
\end{equation}
Here, the symbol $\alltrace$ refers to traces on spinor, flavor
and color indices, and $S(p)$ is the single quark propagator
\begin{equation}
S(p)=\frac{i}{p\hspace{-0.5em}\raisebox{-0.2ex}{/}-m+i\epsilon }=i\frac{p%
\hspace{-0.5em}\raisebox{-0.2ex}{/}+m}{p^{2}-m^{2}+i\epsilon },
\label{qprop}
\end{equation}
of a quark with constituent mass $m$, which is generated from the bare mass $%
\mu _{0}$ via the gap equation Eq.~(\ref{masaquark})
~\cite{Vogl:1991qt,Klevansky:1992qe}. The quantity $\Phi (P,p)$ is
the momentum space image of the Bethe-Salpeter amplitude of a
bound state with total four-momentum $P$, while $p$ is the
four-momentum of one of the constituents. The solution of
Eq.~(\ref{NJLBS}) is rather trivial since the integral on the
right hand side is just a constant, so we can write
\begin{equation}
\vec{\Phi}(P,p)=ig_{\pi qq}S(p)\vec{\tau}\gamma _{5}S(p-P)=-ig_{\pi qq}\frac{%
(p\hspace{-0.5em}\raisebox{-0.2ex}{/}+m)\vec{\tau}\gamma _{5}(p\hspace{-0.5em%
}\raisebox{-0.2ex}{/}-P\hspace{-0.6em}\raisebox{0.2ex}{/}+m)}{%
(p^{2}-m^{2}+i\epsilon )[(P-p)^{2}-m^{2}+i\epsilon ]},
\label{NJLsol}
\end{equation}%
where $g_{\pi qq}$ is the quark-pion coupling constant (given in Eq. (\ref%
{gpqq})) which can be determined from the usual Bethe-Salpeter
normalization
condition \cite{Klevansky:1992qe}. Reinserting this solution into Eq.~(\ref%
{NJLBS}) gives a self-consistency condition, (given in Eq.
(\ref{masapi})), which determines the mass of the ground state as
a function of the coupling constant. Note that in the chiral
limit, when $P^{2}=0$ and $\mu _{0}=0$, Eq.~(\ref{masapi}) is
nothing but the gap equation (\ref{masaquark}), providing some
evidence for the self-consistency of the procedure.

The NJL model has been investigated quite extensively in different
domains
of physics with rather impressive success (see refs.~\cite%
{Vogl:1991qt,Klevansky:1992qe} for reviews), it is therefore
natural to test its predictions for GPDs of the pion. Similar
studies have been carried out
recently~\cite%
{Polyakov:1999gs,Anikin:2000th,Vogt:2001if,Choi:2001fc,Kisslinger:2001gw}
with different models. One drawback of the NJL model is of course
its non-renormalizability, which makes it useful only as an
effective, low-energy model that, however, may be regarded as a
non-linear realization of the QCD Lagrangian. Numerical results
therefore usually depend on the regularization scheme employed to
deal with the divergent integrals. As thoroughly discussed in
refs.~\cite{Davidson:2001cc,Davidson:1995uv}, a suitable
regularization method has to satisfy a certain number of
requirements. The method that was found to be most suitable was a
Pauli-Villars with two subtractions, this is the one that we shall
adopt, as
outlined in appendix~\ref{appa}.

\subsection{Generalized parton distribution}

\begin{figure}[tbp]
\begin{center}
\includegraphics[width=20em]{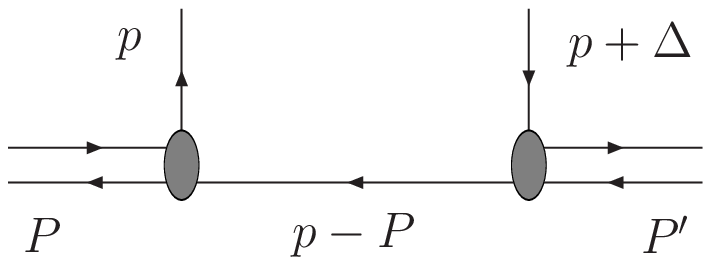} \hspace{4em} %
\includegraphics[width=20em]{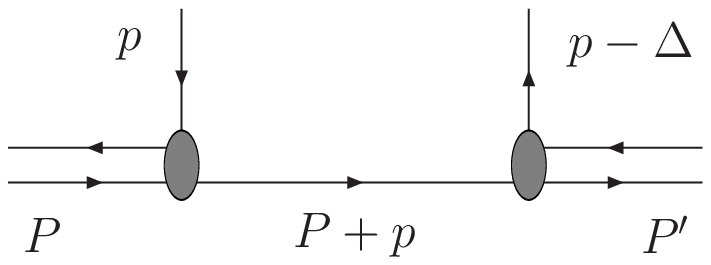}
\end{center}
\caption{Diagrams contributing to the GPD in the Nambu--Jona-Lasinio
Model. }
\label{fighandbag}
\end{figure}

\noindent
We are interested in calculating the diagrams of Fig.~\ref{fighandbag}
in the Nambu--Jona-Lasinio Model. A general
expression for the matrix element of the bi-local $u$-quark current
in the Bethe-Salpeter approach is
\begin{eqnarray}
\Braket{P' | \bar{\Psi}_{u}(x')\gamma_{\mu}\Psi_{u}(x) | P} &=&\int
d^{4}x_{2}\alltrace
\left\{ \bar{\Phi}_{P^{\prime }}^{i}(x^{\prime},x_{2})
\frac{1}{2}(1+\tau _{3}) \gamma _{\mu }
\Big[ \Phi _{P}^{i}(x,x_{2})
(i\overset{\leftarrow }{\slashed{\partial}}{}^{(2)}- m_{2})\Big] \right\}
\notag \\
&&\qquad \qquad +\int d^{4}x_{1}\alltrace
\left\{ \bar{\Phi}_{P^{\prime }}^{i}(x_{1},x)
(i\overset{\rightarrow }{\slashed{\partial}}{}^{(1)}-m_{1})
\Phi _{P}^{i}(x_{1},x^{\prime })\frac{1}{2}%
( 1+\tau _{3}) \gamma _{\mu }\right\},
\label{bso.01}
\end{eqnarray}
where $i$ is the isospin index. The first (second) term in this
expression corresponds to the contribution of the first (second)
constituent of the system, and indices $1$ and $2$ refer to
coordinates or operators related to particles $1$ and $2$, bound in
the meson. In terms of momentum variables we have
\begin{equation}
\vec{\Phi}_{P}\left( x_{1},x_{2}\right) =e^{-iP\cdot X}\int \frac{d^{4}k}{%
\left( 2\pi \right) ^{4}}e^{-ikr}\vec{\Phi}\left( k,P\right) ,
\label{bso.02}
\end{equation}
with the center-of-mass and relative coordinates defined by
\begin{equation}
X=\mu _{1}x_{1}+\mu _{2}x_{2},\qquad r=x_{1}-x_{2},\qquad \mu _{1,2}=\frac{%
m_{1,2}}{m_{1}+m_{2}},  \label{bso.03}
\end{equation}
and $P=p_{1}+p_{2}$, $k=\mu _{2}p_{1}-\mu _{1}p_{2}$ the total and relative
four-momentum of the system.

We shall concentrate on the $u$-quark parton distribution in a $\pi^{+}$
meson, for which we have to consider the first diagram of Fig.~\ref{fighandbag},
with $p_{1}=p$, $p_{2}=P-p$. The second diagram of Fig.~\ref{fighandbag}
contributes to the $d$-quark parton distribution
with $p_{1}=-p$, $p_{2}=P+p$. Inserting Eq.~(\ref{bso.02}) and taking
$m_{1}=m_{2}=m$, the diagrams of Fig.~\ref{fighandbag} give the following
contribution to the GPDs:
\begin{eqnarray}
\lefteqn{\frac{1}{2}\left. \int \frac{dz^{-}}{2\pi }e^{ixP^{+}z^{-}}\Braket{%
P' | \bar{\Psi}_{u}(0) \gamma^+\Psi_{u} (z) | P}\right\vert _{z^{+}=z^{\perp
}=0}=\mathcal H\left( x,\zeta ,t\right) =}  \notag \\
&&\qquad =\frac{1}{2}\int \frac{d^{4}p}{(2\pi )^{4}}\delta \left(
x-p^{+}/P^{+}\right) \alltrace\left\{ \rule[-1ex]{0em}{3ex}\bar{%
\Phi}^{\pi ^{+}}\left( p+P^{\prime }-P,P^{\prime }\right) \frac{1}{2}\left(
1+\tau _{3}\right) \gamma ^{+}\Phi ^{\pi ^{+}}(p,P)
\left( \slashed{p}-\slashed{P}-m\right) \right\},
\label{bso.04}
\end{eqnarray}
where we have switched back to particle momenta. With the
Bethe-Salpeter amplitude of Eq.~(\ref{NJLsol}) we obtain an
expression for the GPD in the Nambu--Jona-Lasinio model that is
similar to Eq.~(\ref{spdfbsp}):
\begin{equation}
\mathcal{H}(x,\zeta ,t)=i4N_{c}g_{\pi qq}^{2}\,\int \frac{d^{4}p}{(2\pi )^{4}%
}\delta (x-p^{+}/P^{+})\,\frac{(x+1-\zeta )(p^{2}-m^{2})+p\cdot \Delta
-xP\cdot \Delta -(2x-\zeta )p\cdot P}{(p^{2}-m^{2}+i\epsilon )\left[
(p+\Delta )^{2}-m^{2}+i\epsilon \right] \left[ (P-p)^{2}-m^{2}+i\epsilon %
\right] }.  \label{spdfnjl}
\end{equation}
The $p^{-}$ integral in Eq.~(\ref{spdfnjl}) is evaluated by the usual
residue calculus. Due to the pole structure of the integrand we obtain two
contributions, the first one in the region $\zeta <x<1$, corresponding to
the quark contribution and the second in the region $0<x<\zeta $,
corresponding to a quark--anti-quark contribution. The second diagram of
Fig.~\ref{fighandbag} will give an anti-quark contribution in the region $%
\zeta -1<x<0$, and a quark-antiquark contribution in the region $0<x<\zeta $%
, just like the first diagram. In case of a $\pi ^{+}$ meson, the first
diagram gives the $u$-quark distribution, while the second one gives the $%
\bar{d}$-quark distribution, but due to isospin symmetry, both distributions
are related by $\mathcal{H}_{u}(x,\zeta ,t)=-\mathcal{H}_{\bar{d}}(\zeta
-x,\zeta ,t)$. Concentrating only on the $u$-quark distribution and
employing a Pauli-Villars regularization, we get for $\zeta<x<1 $:
\begin{eqnarray}
\left. \mathcal{H}(x,\zeta ,t)\rule[-1ex]{0em}{2ex}\right\vert _{\zeta<x<1 }
&=&\frac{N_{c}g_{\pi qq}^{2}}{4\pi ^{2}}\sum_{j=0}^{2}c_{j}\Bigg\{-\log
\frac{m_{j}^{2}}{m^{2}}-\frac{1}{2}\log \frac{m_{j}^{2}-\bar{x}M^{2}}{%
m_{j}^{2}}-\frac{1}{2}\log \frac{m_{j}^{2}-\bar{y}M^{2}}{m_{j}^{2}}  \notag
\label{njlxgtzspd} \\
&&\hspace{2cm}+\frac{(2x-\zeta )M^{2}+(1-x)t}{2}\tilde{I}_{3}(m_{j},x,\zeta
,t)\Bigg\},
\end{eqnarray}
with the abbreviations $\bar{x}=x(1-x)$, $\bar{y}=(1-x)(x-\zeta
)/(1-\zeta )^{2}$, while $\tilde{I}_{3}(m,x,\zeta ,t)$ is given by
Eq.~(\ref{regul4}) with $m = m_{0}$.

\begin{figure}[tbp]
\begin{center}
\includegraphics[width=20em]{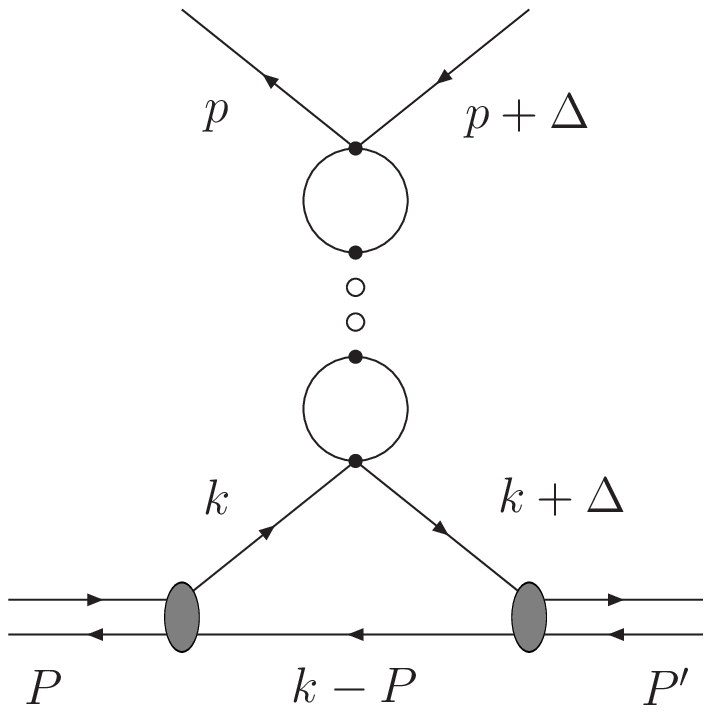} \hspace{5em}
\includegraphics[width=20em]{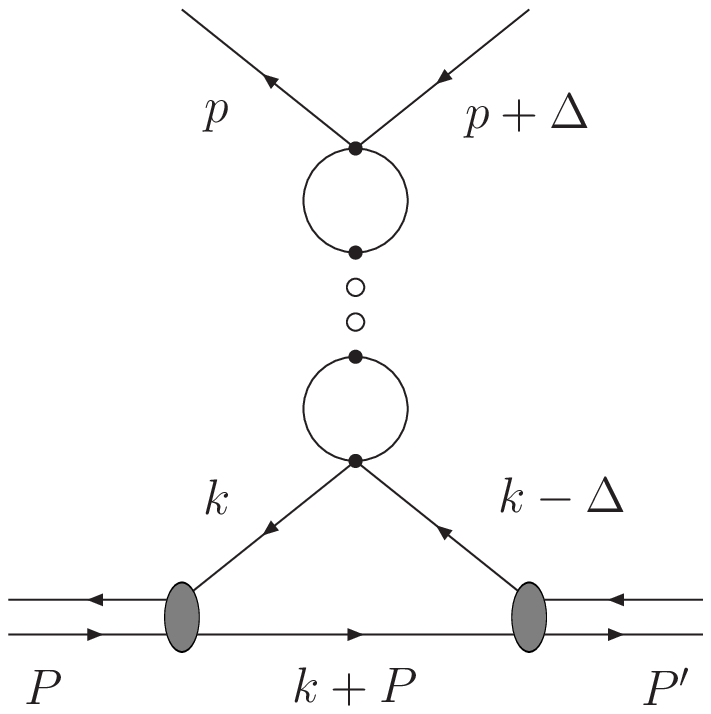}
\end{center}
\caption{Diagrams contributing to the GPDs in the
Nambu--Jona-Lasinio model coming from the "$\sigma$" coupling to
the two photons. They contribute in the region $0<x<\zeta$.}
\label{figsigmadiagram}
\end{figure}

Turning our attention to the non-valence region, $0<x<\zeta$,
we find a first contribution arising from the diagram of
Fig.~\ref{fighandbag}, which is given by
\begin{eqnarray}
\left. \mathcal{H}^{a}(x,\zeta ,t)\rule[-1ex]{0em}{2ex}\right\vert
_{0<x<\zeta } &=&\frac{N_{c}g_{\pi qq}^{2}}{4\pi ^{2}}\sum_{j=0}^{2}c_{j}%
\Bigg\{-\frac{x}{\zeta }\log \frac{m_{j}^{2}}{m^{2}}-\frac{1}{2}\log \frac{%
m_{j}^{2}-\bar{x}M^{2}}{m_{j}^{2}}-\frac{2x/\zeta -1}{2}\log \frac{m_{j}^{2}-%
\bar{y}t}{m_{j}^{2}}  \notag \\
&&\hspace{2cm}+\frac{(2x-\zeta )M^{2}+(1-x)t}{2}\tilde{I}_{3}(m_{j},x,\zeta
,t)\Bigg\},  \label{njlxltzspda}
\end{eqnarray}
where now $\bar{y}=x(\zeta -x)/\zeta ^{2}$.
In the region $0<x<\zeta$, there is a second contribution coming from the
re-scattering diagrams of Fig.~\ref{figsigmadiagram}. They correspond to the
coupling of the two photons in a channel with the quantum numbers of the
$\sigma$ meson. Although not apparent, their contribution is of the same
order as the one from the diagrams in Fig.~\ref{fighandbag} (see
appendix~\ref{appsigmadiagram} for details of the calculation). The
result for the sum of these two diagrams is
\begin{equation}
\left. \mathcal{H}^{b}(x,\zeta ,t)\rule[-1ex]{0em}{2ex}\right\vert
_{0<x<\zeta }=\frac{N_{c}g_{\pi qq}^{2}}{4\pi ^{2}}\left( 1-\frac{2x}{\zeta }%
\right) C\left( t\right) \sum_{j=0}^{2}c_{j}\left[ -\ln \frac{m_{j}^{2}}{%
m^{2}}-\ln \frac{m_{j}^{2}-t\bar{y}}{m_{j}^{2}}\right],
\label{njlxltzspdb}
\end{equation}
with
\begin{equation}
C\left( t\right) =m^{2}\frac{\left[ \left( t-2M^{2}\right) \,I_{3}\left(
P,P^{\prime }\right) +2I_{2}\left( P-P^{\prime }\right) \right] }{%
M^{2}\,I_{2}\left( M^{2}\right) +\left( 4m^{2}-t\right)
I_{2}\left( P-P^{\prime }\right) },
\label{C(t)}
\end{equation}
and $\bar{y}=x(\zeta -x)/\zeta ^{2}.$

In the exact chiral limit, $M = m_{\pi }=0$ and $ m_{\sigma }=2m,$
Eq.~(\ref{C(t)}) becomes
\begin{equation}
C\left( t\right) =\frac{m^{2}}{\left( 4m^{2}-t\right) }\left[ 2+t\,\frac{%
I_{3}\left( P,P^{\prime }\right) }{I_{2}\left( P-P^{\prime
}\right) }\right],
\end{equation}
where the $\sigma$ propagator appears in an explicit way. This
propagator is also present in Eq.~(\ref{C(t)}), but it is
not so apparent due to the more complicated structure of the
equation. Thus we can speak effectively of a coupling of the two
photons to the $\sigma$ state, and this contribution is a
requirement of chiral symmetry.

The total value for the GPD in the region $0<x<\zeta $ is obtained
by summing the two,
\begin{equation}
\left. \mathcal{H}(x,\zeta ,t)\rule[-1ex]{0em}{2ex}\right\vert
_{0<x<\zeta
}=\left. \mathcal{H}^{a}(x,\zeta ,t)\rule[-1ex]{0em}{2ex}\right\vert
_{0<x<\zeta }+\left. \mathcal{H}^{b}(x,\zeta ,t)\rule[-1ex]{0em}{2ex}%
\right\vert _{0<x<\zeta }.
\label{njlxltzspd}
\end{equation}
We note that the first three terms in curly brackets of Eq.~(\ref{njlxgtzspd})
give contributions that are independent of the
momentum transfer $t$. The first term, in particular, gives an
overall constant, independent of $x$, $\zeta $ and $t$, which does
not vanish for $x=1$. From Eq.~(\ref{njlxgtzspd}) we get
the distribution function at zero momentum transfer,
\begin{equation}
q(x)=\mathcal{H}(x,0,0)=\frac{N_{c}g_{\pi qq}^{2}}{4\pi ^{2}}%
\sum_{j=0}^{2}c_{j}\Bigg\{-\log \frac{m_{j}^{2}}{m^{2}}-\log \frac{m_{j}^{2}-%
\bar{x}M^{2}}{m_{j}^{2}}+\frac{m_{j}^{2}}{m_{j}^{2}-\bar{x}M^{2}}\Bigg\},
\label{njlFx00}
\end{equation}
which is consistent with the normalization condition.

\begin{figure}[tbp]
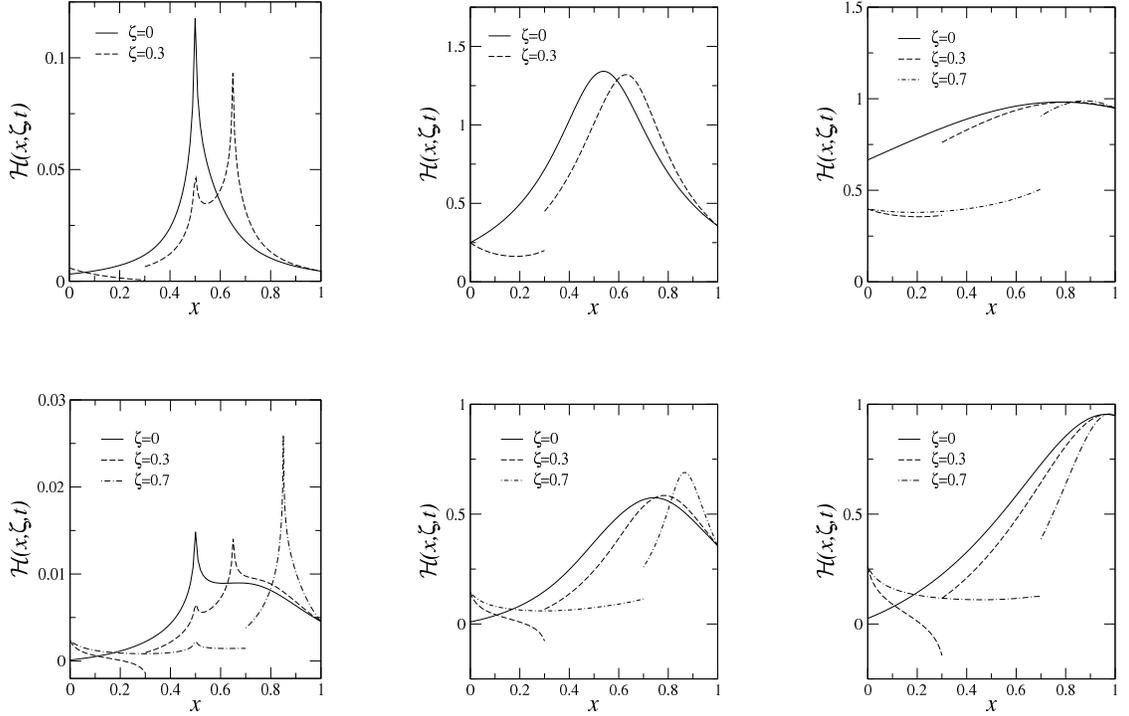

\begin{center}
\includegraphics*[width=13em]{Paper_Def_NJL_wea_01.eps} \hspace{1cm}
\includegraphics*[width=12.5em]{Paper_Def_NJL_450_01.eps} \hspace{1cm}
\includegraphics*[width=12.5em]{Paper_Def_NJL_140_01.eps} \\
\vspace{1.0cm}
\includegraphics*[width=13em]{Paper_Def_NJL_wea_10.eps} \hspace{1cm}
\includegraphics*[width=12.5em]{Paper_Def_NJL_450_10.eps} \hspace{1cm}
\includegraphics*[width=12.5em]{Paper_Def_NJL_140_10.eps}
\end{center}
\caption{Generalized parton distributions in the
Nambu--Jona-Lasinio model for different values of the bound state mass
$M$ and momentum transfers $t$. The top row gives results for
$t = -0.1$ GeV$^2$, while on the bottom for $ t = -1$ GeV$^2$. We
describe from left to right the scenarios discussed in the text, i.e., from weak
binding ($M\approx 2m$, left column), to strong binding (middle column)
to the chiral limit ($M=0$, right column). Note that the sum rule of
Eq.~(\protect\ref{relat}) is exactly satisfied for each graph.}
\label{fignjl}
\end{figure}

In Fig.~\ref{fignjl} we give some examples of GPDs in the NJL
model for different values of the binding energy and momentum
transfers. We distinguish in here three scenarios: i) the weak
binding scenario $M\approx 2m$ (up to small binding energies of
electromagnetic magnitude) ; ii) the deep binding scenario, with
binding energies in the tens of MeV range, which would correspond
physically to a strongly bound state, i.e. like the $\rho$ meson;
iii) the Goldstone boson scenario, which we take close to the
chiral limit, $M = m_\pi\approx 140$ MeV. This classification is
very natural as the behavior of the GPDs in the figures shows. The
Goldstone boson scenario, with its flat GPDs, did not appear in
scalar Electrodynamics and it shows how spontaneous chiral
symmetry breaking appears in the GPDs.

For zero binding, there is no spontaneous symmetry breaking, thus
no pion, and the constituent quarks are quasi free. For weak
binding, we expect for the GPDs  a similar behavior to the one
discussed in scalar Electrodynamics. This
similarity can be seen by comparing the corresponding graphs of
Figs.~\ref{figsed} and~\ref{fignjl}.
In the weak binding limit we note again the
appearance of two peaks at $x=1/2$ and $x=(1+\zeta )/2$, which go over into
$\delta$-functions in the exact limit $M=2m$, just as in the
scalar case, see Eq.~(\ref{deltaGPD}).
As soon as there is binding
the GPDs differ more and more from the free ones as we approach
the chiral $M=m_{\pi }$ limit with massive constituent quarks.

A few properties of the final result may be recognized from the
analytic expressions of the GPDs. First, the total GPDs are
discontinuous at $x=\zeta$. The discontinuity arises from the
diagrams in Fig.~\ref{figsigmadiagram}, since the contribution
from the diagrams in Fig.~\ref{fighandbag} to the GPDs is
continuous, even though the derivative at $x=\zeta$ is not.
Second, we see that $\mathcal{H}(0,\zeta ,t)\neq 0$ for any value
of $\zeta$. Its value, which depends on $t$ and $M$ but not on
$\zeta$, is connected to the discontinuity by:
\begin{equation}
\mathcal{H}\left( 0,\zeta ,t\right) =\mathcal{H}\left( \zeta
_{+},\zeta ,t\right) -\mathcal{H}\left( \zeta _{-},\zeta ,t\right)
\neq 0 \label{H(0)=Discontinuidad}
\end{equation}
Third, as noted above, $\mathcal{H}(1,\zeta ,t)\not=0$. Its value
is independent of $t$ and $\zeta ,$ and only depends on the bound
state mass $M$ via the coupling constant $g_{\pi qq}$. In
particular, in the chiral limit, when $M=0,$ the value of
$\mathcal{H}(x,\zeta ,t)$ at $x=1$ is simply given by:
\begin{equation}
\mathcal{H}(1,\zeta ,t)=1.  \label{relatpi}
\end{equation}
Furthermore, the distribution functions become relatively more
concentrated around $x=1$ for large $|t|$, which is just a
consequence of the constant value of the distribution function at
$x=1$. Finally, in spite of these peculiarities, we find again
that Eq.~(\ref{relat}) is always exactly satisfied, as we checked
numerically.

The polynomiality condition in this case takes a very simple form
in the chiral limit. When $M = 0$, the following relation holds:
\begin{equation}
F_{N}(\zeta ,0)=\frac{1}{N}\left( 1-\frac{1}{2}\zeta ^{N}\right) ,
\label{njlpoly}
\end{equation}
i.e., at zero momentum transfer, the polynomiality condition is
fulfilled. Again, for $N = 1,$ Eq.~(\ref{njlpoly}) reproduces
Eq.~(\ref{relat}) in the case $t=0$. For $t\neq 0$ and $%
M^{2}\neq 0$ we have shown, by numerical integration, that the
polynomiality condition also holds. Further details are given in
appendix~\ref{polynomiality}.

The fact that the polynomiality condition holds both for SED and
NJL in the present scheme should not be a surprise since both
models arise from a field theoretic description which preserves
Lorentz symmetry, parity, time-reversal invariance and
hermiticity~\cite{Ji:1998pc}.

Finally we note that in the chiral limit, $M = 0$ and at zero
momentum transfer, $t = 0,$\ we reproduce the analytic results of
ref.~\cite{Polyakov:1999gs}, namely:
\begin{equation}
\mathcal{H}(x,\zeta ,0)=\frac{1}{2}\theta \left( \zeta -x\right) \,\theta
\left( x\right) +\theta \left( x-\zeta \right) \,\theta \left( 1-x\right) ,
\label{M0t0}
\end{equation}
so in particular, $\mathcal{H}(x,1,0)=1/2$ and $\mathcal{H}(x,0,0)
= 1$. It may be checked from Eq.~(\ref{spdfnjl}) that this result
is completely independent of any regularization procedure. This form may be
compared with the corresponding result in the scalar model, see
Eq.~(\ref{M0t0sca}). Eq.~(\ref{M0t0}) also implies that in the
chiral limit we get a quark distribution function that is equal to
unity. This coincides with the results obtained in
refs.~\cite{Weigel:1999pc,Davidson:1995uv,Shigetani:1993dx}.

\subsection{Isospin decomposition}

Following the idea of ref.~\cite{Polyakov:1998ze} we introduce an
isospin decomposition for the GPDs of the pion\footnote{%
Note that our notation is slightly different from the cited
reference.}
\begin{equation}
\int \frac{dz^{-}}{2\pi }e^{ixP^{+}z^{-}}\left.
\Braket{\pi^{a}(P^{\prime})|\bar{\Psi}_{f^\prime}(0) \gamma^+\Psi_{f}(z)|\pi ^{b}(P)}
\right\vert _{z^{+}=z^{\perp }=0}
=\delta ^{ab}\delta _{f\,f^{\prime }}\mathcal{H}^{I=0}\left( x,\zeta
,t\right) +\varepsilon ^{abc}\tau _{f\,f^{\prime }}^{c}\mathcal{H}%
^{I=1}\left( x,\zeta ,t\right).
\label{IsosDec}
\end{equation}
In order to extract the two isospin components we need to
calculate the $u$-quark GPD for the $\pi ^{0}$. We have
contributions coming from both diagrams of Fig.~\ref{fighandbag}
and from those of Fig.~\ref{figsigmadiagram}, leading to
\begin{equation}
\mathcal{H}_{u}^{\pi ^{0}}( x,\zeta ,t) =\frac{1}{2}\big[
\mathcal{H}\left( x,\zeta ,t\right) -\mathcal{H}\left( \zeta -x,\zeta
,t\right) \big],
\label{GPDpi0}
\end{equation}
where $\mathcal{H}\left( x,\zeta ,t\right) $ is the $u$-quark GPD for the
$\pi ^{+}$ defined in Eqs.~(\ref{njlxgtzspd}) and~(\ref{njlxltzspd}).
Here $x$ ranges from $\zeta-1$ to $1,$ as has been explained below
Eq.~(\ref{spdfnjl}). The GPDs of the $\pi^{+}$ and $\pi^{0}$ allow to extract
the isospin components
\begin{equation}
\begin{array}{lll}
\mathcal{H}^{I=0}\left( x,\zeta ,t\right) & =2\mathcal{H}_{u}^{\pi
^{0}}\left( x,\zeta ,t\right) & =\mathcal{H}\left( x,\zeta ,t\right) -%
\mathcal{H}\left( \zeta -x,\zeta ,t\right) \\
\mathcal{H}^{I=1}\left( x,\zeta ,t\right) & =2\mathcal{H}^{\pi ^{+}}\left(
x,\zeta ,t\right) -2\mathcal{H}_{u}^{\pi ^{0}}\left( x,\zeta ,t\right) & =%
\mathcal{H}\left( x,\zeta ,t\right) +\mathcal{H}\left( \zeta -x,\zeta
,t\right).%
\end{array}%
\end{equation}
Using Eq.~(\ref{M0t0}) we obtain simple expressions for the two
isospin GPDs in the $t = 0,\,m_{\pi } = 0$ case:
\begin{equation}
\mathcal{H}^{I=0}\left( x,\zeta ,t\right) =\left\{
\begin{array}{rcr}
-1, && \zeta -1<x<0 \\
0, && 0<x<\zeta \\
1, && \zeta <x<1%
\end{array}%
\right.
\qquad ; \qquad
\mathcal{H}^{I=1}\left( x,\zeta ,t\right) =1,\qquad \zeta -1<x<1.
\end{equation}
We observe that the predictions of ref.~\cite{Polyakov:1998ze},
regarding the isospin components in the limit $m_{\pi }=0$, are
satisfied. The relation between $\mathcal{H}^{I=1}\left( x,\zeta
=1,t=0\right) $ and the wave function of the pion, $\varphi _{\pi
}(x)$, and  $\mathcal{H}^{I=0}\left( x,\zeta =1,t=0\right) =0$
hold. Note moreover that our pion wave function, in this limit, is
$\varphi _{\pi }(x) = 1$. The $m_{\pi } = 0,$ $\zeta =1$ and $t =
0$ case corresponds to a process in which the initial and final
pion momenta are $P=\left( P^{+},\vec{0}^{\perp },0\right) $ and
$P^{\prime }=\left( 0,\vec{0}^{\perp },0\right) $, respectively.
For zero pion mass this process is allowed only as a limit. In the
NJL model we can look for deviations from the exact chiral limit.

\subsection{Electromagnetic form factor}

\noindent In order to understand better the above results, we may want to
calculate the electromagnetic form factor in the NJL model which is given by
the $x$-integrated parton distribution, see Eq.~(\ref{relat}). This has been
done already in several
works~\cite{Bernard:1988bx,Klevansky:1992qe,Schulze:1994fy},
but we want to investigate
the relation with the present results. Calculating again the current we
find, from the first diagram of Fig.~\ref{fig1}, for the form factor and for
the GPDs:
\begin{equation}
(P+P^{\prime })^{\mu }\,F(Q^{2})=8iN_{c}g_{\pi qq}^{2}\int \frac{d^{4}p}{%
(2\pi )^{4}}\frac{(p^{\mu }+P^{\mu }+\Delta ^{\mu })(p^{2}-m^{2}+i\epsilon
)+p\cdot \Delta \,P^{\mu }-P\cdot \Delta \,p^{\mu }-(2p^{\mu }+\Delta ^{\mu
})\,p\cdot P}{(p^{2}-m^{2}+i\epsilon )\left[ (p+\Delta )^{2}-m^{2}+i\epsilon %
\right] \left[ (P-p)^{2}-m^{2}+i\epsilon \right] }.
\label{njlff1}
\end{equation}
Rewriting the denominator under the integral and using some relations of
appendix~\ref{appb}, we arrive at
\begin{equation}
F(Q^{2})=i2N_{c}g_{\pi qq}^{2}\frac{(Q^{2}/2+M^{2})I_{2}(\Delta
)+M^{2}I_{2}(P)-M^{4}I_{3}(\Delta ,-P)}{Q^{2}/4+M^{2}}.
\label{njlff3}
\end{equation}
We checked numerically (by applying the same Pauli-Villars regularization
method) that this form factor is exactly reproduced by integrating our
result for the GPD over $x$. This means that the non-vanishing distribution
functions (at $x=1$) are implicitly present in the result of
Eq.~(\ref{njlff3}). Note that in the chiral limit, $M^{2}=0$, we simply have
\begin{equation}
F(Q^{2})=4iN_{c}g_{\pi qq}^{2}I_{2}(\Delta )=1-R(Q^{2})/I_{2}(0),
\label{njlff4}
\end{equation}%
where for small $Q^{2}$, $R(Q^{2})$ is given by
\begin{equation}
R(Q^{2})=\frac{i}{16\pi ^{2}}\frac{Q^{2}}{6m^{2}},
\label{njlff6}
\end{equation}
so together with the relation $f_{\pi }^{2}=-12im^{2}I_{2}(0)$, we can
evaluate the root-mean squared radius of the pion in this model via
$\left\langle r_{\pi }\right\rangle ^{2}=-6\,\left.\partial F(Q^{2})/\partial
Q^{2}\right|_{Q^2=0}=3f_{\pi }^{-2}/(4\pi ^{2})$, which is the same as in
ref.~\cite{Klevansky:1992qe}. With $f_{\pi }=93$ MeV, we find the numerical value of $%
\left\langle r_{\pi }\right\rangle ^{2}=(0.585\,\mathrm{fm})^{2}$, to be
compared to the experimental value of $(0.66\,\mathrm{fm})^{2}$. Note,
however, that Eq.~(\ref{njlff6}) is valid only in the limit when the cut-off
goes to infinity, for the finite cut-off as specified in appendix~\ref{appa}%
, the root mean squared radius gets multiplied by a factor $\sim 0.89$.

\section{Discussion}

\noindent
The results obtained in the scalar Electrodynamics model
show a perfect realization of all wishful ingredients. The
calculation is exact, finite and satisfies all the desired
properties, i.e., the GPDs have the correct support and vanish at
the boundary regions, while the sum rule and the polynomiality
condition are exactly verified. From the physical point of view,
the GPDs show a realization in terms of quasi-free constituents in
the weak binding limit. As the binding increases one is confronted
with the dynamics as derivable from a non trivial momentum
distribution determined by the corresponding Bethe-Salpeter
amplitude, a feature also appearing in other model
calculations~\cite{Scopetta:2002xq}. The physical effect
associated with $t$ and $\zeta$ is naturally represented in the
GPDs. The variable $t$ tends to push the constituent distribution
towards higher values of $x$, which corresponds to an input of
momentum transfer into the system, while the variable $\zeta$
incorporates the description of virtual pairs. Unfortunately this
model is not very realistic for the pion (perhaps it might be
better fit for a description of the nucleon) and our results
represent only qualitative features of how the dynamics might
influence the distributions.

An equivalent model was considered in
refs.~\cite{Tiburzi:2001ta,Tiburzi:2001je}, but the emphasis was put on the light-front
quantization method. The results were calculated only for $x>\zeta$, while
the region $x<\zeta$ was explored by an analytic continuation of the vertex
function. Due to the approximations used in their approach the continuation
is not perfect and therefore sum rules like the one appearing in
Eq.~(\ref{relat}) are explicitly violated.

For the NJL model the calculation requires regularization. The latter
certainly influences the results, as has been discussed in
ref.~\cite{Davidson:2001cc}, where the Pauli-Villars method was compared to the one of
Brodsky-Lepage with different results. The Pauli-Villars method is
compatible with all the symmetry requirements~\cite{Weigel:1999pc}, which is
the reason for our choice. A caveat that should be emphasized is that the
model does not only contain the dynamics expressed in the Lagrangian, but
also the one derived from the regularization procedure.

\begin{figure}[tp]
\begin{center}
\includegraphics[width=15em]{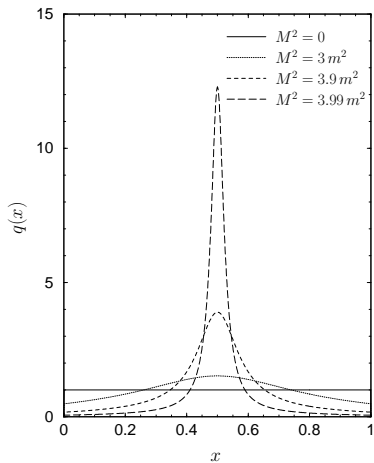}\hspace{4em}
\end{center}
\caption{Quark distribution function $q(x)={(x,0,0)}$ in the
Nambu--Jona-Lasinio model. The constituent mass is kept fixed at $m =
241\,\mathrm{MeV}$. } \label{figmass}
\end{figure}

The NJL calculation retains some nice properties, in particular it
preserves the sum rule and the polynomiality condition. Physically
it is also very appealing since we can distinguish features
associated with the weak and deep binding regimes. In
Fig.~\ref{figmass} we show  the variation of the quark
distribution function with the binding energy. As the binding
energy increases we change softly from a delta type behavior, at
zero binding, to a constant behavior in the strong binding regime.
This figure also illustrates nicely the phase transition
associated with the spontaneous breaking of chiral symmetry.
Suppose we keep the mass of the bound state fixed at $M^{2}=0$ and
consider the variation of $q(x)$ with the constituent mass $m$. It
is clear that we will always have $q(x)=1$ for any value of $m$,
except when $m = 0$, where the distribution function
changes discontinuously to $q(x)=\delta (x-1/2)$. The effects of $t$ and $%
\zeta $ described for scalar Electrodynamics persist, as can be seen in the
graph on the right hand side of this figure, i.e., $t$ pushes the
distribution to higher values of $x$ and $\zeta $ introduces virtual pairs
into the description of the system.

Unexpected  results of the calculation are the non-vanishing of
the GPDs at the boundary, i.e., $\mathcal{H}(x,\zeta ,t)\not=0$
for $x=1$ and $x=0$. This feature has nothing to do with what is
usually called the support problem. The latter is characterized by
a non-vanishing distribution function outside the physical region,
i.e. $\mathcal{H}(x,\zeta ,t)\not=0$ for $x<\zeta -1$ or $x>1$,
while in our calculations, the GPDs explicitly vanish there. It is
therefore really a discontinuity that we encounter at the physical
boundary.

For the quark parton distribution $q(x)=\mathcal{H}(x,0,0)$, this
peculiarity was already noted in
refs.~\cite{Weigel:1999pc,Davidson:1995uv,Shigetani:1993dx} and it is
obviously present in an implicit way in the results of electromagnetic form
factor calculations~\cite{Bernard:1988bx,Klevansky:1992qe,Schulze:1994fy}.
It showed up already in several model
calculations~\cite{Polyakov:1999gs,Praszalowicz:2002ct,RuizArriola:2002bp,Dorokhov:2000gu,Anikin:1999cx}.

Another peculiarity of the calculation is the discontinuity at the
$ x=\zeta $. In the NJL model, the value of the GPD at $x=0$ and
the discontinuity at $x=\zeta$ are related, as we have shown in
Eq.~(\ref{H(0)=Discontinuidad}), for any value of $M$, $t$ and
$\zeta $. The value of $\mathcal{H}\left( 0,\zeta ,t\right) $
decreases as $t$ increases or as the binding energy decreases. But
the origin of the non zero value of $\mathcal{H}\left( 0,\zeta
,t\right) $  is in the $\sigma$ diagrams depicted in
Fig.~\ref{figsigmadiagram}, which originate from the scalar
four-fermion coupling present in the Lagrangian, Eq.~(\ref{NJLlagem}).

Let us discuss the discontinuity at the boundaries further, within
the NJL description. One might investigate the mass dependence of
the GPDs at a given point $x$. In our calculation we get for $x=1$
the general expression
\begin{equation}
\mathcal{H}(x=1,\zeta ,t)=\frac{F_{0}}{F_{0}-1+\left( \sqrt{\frac{4m^{2}}{%
M^{2}}-1}+\frac{1}{\sqrt{\frac{4m^{2}}{M^{2}}-1}}\right) \arctan {\left(
\frac{1}{\sqrt{\frac{4m^{2}}{M^{2}}-1}}\right) }+\ldots },
\label{Hx1}
\end{equation}
where $F_{0}=16i\pi ^{2}I_{2}(0)$, with $I_{2}(p)$ given by Eq.~(\ref{regul2}%
) and the dots denote small terms of higher order in $M^{2}/\Lambda ^{2}$,
where $\Lambda $ is the cut-off parameter in the Pauli-Villars
regularization, see appendix~\ref{appa}. This equation illustrates the
dependence on the bound state mass, and also the regularization scheme
dependence through $F_{0}$ is apparent. It is clear from Eq.~(\ref{Hx1})
that $\mathcal{H}(x=1,\zeta ,t)$ vanishes for zero binding ($M^{2}=4m^{2}$)
while it is non-vanishing, as soon as the interaction binds the quarks into
a pion, i.e. $M^{2}<4m^{2}$.

Moreover in the NJL model, for a small mass particle like the pion
($M = m_\pi$):
\begin{equation}
\mathcal{H}(x=1,\zeta ,t)=\frac{g_{\pi qq}f_{\pi }}{m_{\pi }}+\mathcal{O}%
(m_{\pi })+\ldots =1+\mathcal{O}(m_{\pi })+\ldots ,  \label{gt}
\end{equation}
where the last step follows from the Goldberger-Treiman relation as a
consequence of spontaneously broken chiral symmetry, and to lowest order in
$m_{\pi}$ this is regularization scheme independent.

The physical interpretation of these features is not completely
clear at present. Let us however conjecture a solution based on an
analysis of known results. The usual idea that the parton
distributions must vanish on the boundary is based in the study of
free parton models. If one of the partons carries all the $+$
component of the momentum, i.e. $x=1$, the other one has $p^{+}=0$
and thus $p^{-}$ goes to infinity if the particle is on shell.
Since this $p^{-}$
enters the propagator, the obvious conclusion is that the parton
distribution must vanish at this boundary. But this simple
description is broken when the particles are off-shell, like those
in a bound system. Our result is therefore not unnatural and does
not contradict any physical intuitive ideas. Moreover, the fact
that the distributions do not vanish at $x=0$ for $\zeta =0$ is
just a consequence of the non-vanishing at $x=1$ and does not seem
to introduce any new conceptual problem.

One could suspect that this feature may originate as an artifact of the
regularization in the NJL model, but this suspicion should be immediately
dropped since it occurs in some limits in a regularization independent way.
Therefore we conclude that there is a deeper physical reason for it, namely
the off-shellness of the bound state quarks and the realization of chiral
symmetry\footnote{Similar features occur in the well defined
't~Hooft model~\protect\cite{'tHooft:1974hx}, a theoretical scheme which is
exact in the large $N_c$ limit.
In the chiral limit, this model produces a parton distribution
$q(x)$ which is equal to 1 for any value of $x$, in a regularization
independent way. But, when chiral symmetry is explicitly broken, the model
of 't~Hooft leads to a parton distribution which vanishes at the end-points,
$q(x=0)=q(x=1)=0$. This different behavior is not surprising since
the dynamics of this model and that of the NJL model are very
different.  The 't~Hooft model is a 1+1 dimensional model, which is
confining, but without a point-like interaction.}.

Up to now we have been discussing the NJL model as a theory in itself.
However the true theory of the strong interactions is QCD and the NJL model
should be interpreted as an effective theory of QCD at low energies. QCD
based arguments led in refs.~\cite{Polyakov:1999gs,Praszalowicz:2002ct} to
define a modified NJL model with a momentum dependent constituent quark mass
(or equivalently, a constituent quark form factor) which, when appropriately
chosen, makes the GPDs vanish at the end points.
In ref.~\cite{RuizArriola:2002bp} the emphasis was placed in the perturbative
aspects, i.e., gluons and sea bremsstrahlung. The QCD evolution
equations~\cite{Muller:1995cn} were used to eliminate the high momentum
components. Certainly both mechanisms should be simultaneously advocated.


\section{Conclusion}

\noindent In this work, we presented a detailed calculation of
generalized parton distribution functions using a covariant Bethe
- Salpeter approach both in scalar Electrodynamics and in the
Nambu--Jona-Lasinio model. No assumptions have gone into the
determination of the Bethe - Salpeter amplitudes or any other
ingredients of the calculation. The only approximations employed
are the ladder approximation for the determination of the Bethe -
Salpeter bound state amplitude (which is, of course, still a fully
covariant object in this case), and in the determination of the
current matrix elements, we have restricted ourselves to the
lowest order diagrams (but we have shown that the next order
correction vanishes exactly). As a result of this procedure, no
important features required by general physical considerations,
like symmetry properties, sum rules, etc., have been violated and
we recover them in our numerical results. For the scalar model,
the calculation evidences all desirable features, we reproduce the
results obtained in similar studies, extend them to other
kinematical regions and find sum rules which were not possible in
the other treatments.

In the case of the NJL model, we found that the realization of
chiral symmetry plays a crucial role in the outcome of the
calculation. In the region $0 < x < \zeta$ chiral symmetry
intervenes explicitly through the contribution from the scalar
channel. In the region  $ \zeta < x < 1$, where the $\sigma$ does not
contribute, it appears implicitly through the consistency
equations which govern the dynamics of the pion and its coupling
to quarks. In the massive case, this reflects
in the fact that the GPDs do not vanish at the boundary of the
kinematic region, i.e., at $x=0$ and $x=1$. Our
detailed analysis associates this feature with the off-shellness of the
constituent quarks, and the specific value there, with
spontaneous chiral symmetry breaking. Moreover the GPDs are
discontinuous at $x=\zeta$, a consequence of the $\sigma$ channel
contribution,i.e., also a requirement of the chiral symmetry
realization. These behaviors have been shown to arise in a
regularization independent way.

In conclusion, we have shown that the SED model has continuous,
well behaved GPDs that vanish at the end-points of the kinematic region,
and bubble diagrams which do not contribute as a
consequence of gauge symmetry. On the other hand, the NJL model has
discontinuous, ill behaved GPDs, that are non-zero at the end-points, and
bubble diagrams which, in the $\sigma$ channel, do contribute as
a consequence of the realization of chiral symmetry.

Before finishing we must recall that our calculation is valid at the
hadronic scale, i.e., at a low momentum renormalization point. Evolution to
higher momenta is necessary to describe deep inelastic scattering data. It
will be interesting to see how the described features of the NJL model will
change under evolution. However the fact that the distributions do not
vanish at the boundary imply the appearance of strong singularities which
render the process non trivial.


\begin{acknowledgments}
This work was supported by the European Commission IHP program
under contract HPRN-CT-2000-00130, MCYT (Spain) under contracts
BFM2001-3563-C02-01 and BMF2001-0262, and Generalitat Valenciana
under contract GV01-216. One of us (V.V.) would like to thank the
Physics Department of Seoul National University, and Prof.
Dong-Pil Min in particular, for the hospitality extended to him
during the last stages of this work. We also want to thank the referees
for their interesting comments which have improved the manuscript
considerably.
\end{acknowledgments}


\appendix

\section{Regularization}

\label{appa}

\noindent We have used the Pauli-Villars regularization in the
Nambu--Jona-Lasinio model in order to render the occurring
integrals finite. This means that for integrals like the ones
defined by Eqs.~(\ref{int1}-\ref{int4}), we make the replacement
\begin{equation}
\int \frac{d^{4}p}{(2\pi )^{4}}f(p;m^{2})\longrightarrow \int \frac{d^{4}p}{%
(2\pi )^{4}}\sum_{j=0}^{2}c_{j}f(p;m_{j}^{2}),  \label{regul}
\end{equation}
with $m_{j}^{2}=m^{2}+j\Lambda ^{2}$, $c_{0}=-c_{1}/2=c_{2}=1$. Following
ref.~\cite{Klevansky:1992qe} we determine the regularization parameters $%
\Lambda $ and $m$ by calculating the pion decay constant and the quark
condensate in the chiral limit via
\begin{equation}
f_{\pi }^{2}=-\frac{3m^{2}}{4\pi ^{2}}\sum_{j=0}^{2}c_{j}\log
(m_{j}^{2}/m^{2}),\qquad \qquad \left\langle \bar{u}u\right\rangle =-\frac{3m%
}{4\pi ^{2}}\sum_{j=0}^{2}c_{j}m_{j}^{2}\log (m_{j}^{2}/m^{2}).
\label{fpiubaru}
\end{equation}
With the conventional values $\left\langle \bar{u}u\right\rangle =-(250\,%
\mathrm{MeV})^{3}$ and $f_{\pi }=93$ MeV, we get $m = 241$ MeV and
$\Lambda =859$ MeV.


\section{Elementary integrals}

\label{appb}

Some of the interesting integrals appearing in the main text are:
\begin{equation}
I_{1}\equiv \int \frac{d^{4}\,k}{(2\pi )^{4}}\frac{1}{k^{2}-m^{2}+i\epsilon }%
,  \label{int1}
\end{equation}
\begin{equation}
I_{2}(p)\equiv \int \frac{d^{4}\,k}{(2\pi )^{4}}\frac{1}{\left(
k^{2}-m^{2}+i\epsilon \right) \left( (k+p)^{2}-m^{2}+i\epsilon \right) },
\label{int2}
\end{equation}
\begin{equation}
I_{3}(p_{1},p_{2})\equiv \int \frac{d^{4}\,k}{(2\pi )^{4}}\frac{1}{\left(
k^{2}-m^{2}+i\epsilon \right) (\left( k+p_{1}\right) ^{2}-m^{2}+i\epsilon
)(\left( k+p_{2}\right) ^{2}-m^{2}+i\epsilon )},  \label{int4}
\end{equation}
From these definitions the following relation may be deduced:
\begin{equation}
\int \frac{d^{4}\,k}{(2\pi )^{4}}\frac{k^{\mu }}{\left(
k^{2}-m^{2}+i\epsilon \right) \left( (k+p)^{2}-m^{2}+i\epsilon \right) }=-%
\frac{1}{2}p^{\mu }I_{2}(p).  \label{int6}
\end{equation}
The scalar model is renormalizable. The NJL model is not. We use in
the latter case the Pauli-Villars regularization procedure and obtain:
\begin{equation}
I_{1}=-\frac{i}{16\pi ^{2}}\sum_{j=0}^{2}c_{j}m_{j}^{2}\log
(m_{j}^{2}/m^{2}),  \label{regul1}
\end{equation}
\begin{equation}
I_{2}(p)=-\frac{i}{16\pi ^{2}}\sum_{j=0}^{2}c_{j}\left\{ \log
(m_{j}^{2}/m^{2})+\sqrt{\frac{p^{2}-4m_{j}^{2}}{p^{2}}}\log \frac{1-\sqrt{%
\frac{p^{2}}{p^{2}-4m_{j}^{2}}}}{1+\sqrt{\frac{p^{2}}{p^{2}-4m_{j}^{2}}}}%
\right\} .  \label{regul2}
\end{equation}
From Eq.~(\ref{int4}) and $p_{1}^{2}=p_{2}^{2}=M^{2}$ we have:
\begin{equation}
I_{3}(p_{1},p_{2})=I_{3}(t=\left( p_{1}-p_{2}\right) ^{2})=-\frac{i}{16\pi
^{2}}\sum_{j=0}^{2}c_{j}\int_{0}^{1}dz\frac{z}{\sqrt{d_{j}}}\ln \left| \frac{%
e_{j}+\sqrt{d_{j}}}{e_{j}-\sqrt{d_{j}}}\right|, \label{intm3}
\end{equation}
with
\begin{equation}
d_{j} = t^{2}z^{4}-4z^{2}t\left( m_{j}^{2}+M^{2}z\left( z-1\right)
\right), \;\;\;\;\;\;  e_{j}  = 2m_{j}^{2}+2M^{2}z\left(
z-1\right) -z^{2}t.
\end{equation}

The self consistency relation determining the constituent mass of
the quark, m, is
\begin{equation}
m = \mu _{0}+8\;i\;g\;N_{c}\;N_{f}\;m\;I_{1}, \label{masaquark}
\end{equation}%
while the equation for the mass of the pion , $P^{2}=M^{2},$ leads
to
\begin{equation}
1=4\;i\;g\;N_{c}\;N_{f}\;\left[ 2I_{1}-P^{2}I_{2}\left( P\right)
\right], \label{masapi}
\end{equation}%
and the pion quark coupling constant , $g_{\pi qq}$, is given by
\begin{equation}
g_{\pi qq}^{2}=\frac{-1}{12i\left( I_{2}\left( M^{2}\right)
+M^{2}\left(
\partial I_{2}\left( p\right) /\partial p^{2}\right) _{p^{2}=M^{2}}\right)
}. \label{gpqq}
\end{equation}

To obtain the results of Eqs.~(\ref{anares},\ref{njlxgtzspd},\ref{njlxltzspda}),
we used the following formulas:
\begin{equation}
\tilde{I}_{2}(P)\equiv -16i\pi ^{2}\int \frac{d^{4}\,k}{(2\pi )^{4}}\frac{%
\delta \left( x-k^{+}/P^{+}\right) }{\left( k^{2}-m^{2}+i\epsilon \right)
\left( (k-P)^{2}-m^{2}+i\epsilon \right) }=\left\{
\begin{array}{ccl}
\displaystyle\sum_{j=0}^{2}c_{j}\log \displaystyle\frac{m_{j}^{2}-\bar{x}%
P^{2}}{m^{2}}, &  & 0<x<1, \\
0 &  & \mathrm{otherwise}.%
\end{array}%
\right.
\label{regul3}
\end{equation}
\renewcommand{\arraystretch}{2.0}
\begin{eqnarray}
\tilde{I}_{3}(m,x,\zeta ,t) &\equiv &-16i\pi ^{2}\int \frac{d^{4}\,k}{(2\pi
)^{4}}\frac{\delta \left( x-k^{+}/P^{+}\right) }{\left(
k^{2}-m^{2}+i\epsilon \right) \left( (k+\Delta )^{2}-m^{2}+i\epsilon \right)
\left( (P-k)^{2}-m^{2}+i\epsilon \right) }=  \notag  \label{regul4} \\
&&  \notag \\
&=&\left\{
\begin{array}{ccl}
\displaystyle\frac{1}{\sqrt{D}}\log \displaystyle\frac{[2(1-\zeta )(\zeta
y-x)-\zeta ]M^{2}-(1-x)t+2m^{2}/y+\sqrt{D}}{[2(1-\zeta )(\zeta y-x)-\zeta
]M^{2}-(1-x)t+2m^{2}/y-\sqrt{D}}, &  & 0<x<1, \\
0 &  & \mathrm{otherwise}.%
\end{array}%
\right.
\end{eqnarray}
In the last integral, $D=\zeta
^{2}M^{2}(M^{2}-4m^{2})+(1-x)^{2}t^{2}+2(1-x)(2x-\zeta
)M^{2}t-4m^{2}(1-\zeta )t$, $\zeta =-\Delta ^{+}/P^{+}$, $M^{2}=P^{2}$, $%
t=\Delta ^{2}$ and
\renewcommand{\arraystretch}{1.0}
\begin{equation}
y=\left\{
\begin{array}{ccl}
x/\zeta , &  & 0<x<\zeta , \\
(1-x)/(1-\zeta ), &  & \zeta <x<1.%
\end{array}%
\right.  \label{lastone}
\end{equation}


\section{Contribution of the $\sigma$ - diagrams}
\label{appsigmadiagram}

\noindent
The contribution of the diagrams shown in  Fig.~\ref{figsigmadiagram}
can be calculated as $\mathcal{H}^{b}\left( x,\zeta ,t\right) =A\cdot B$, with
\begin{equation}
A=\frac{1}{2}\int \frac{d^{4}p}{\left( 2\pi \right) ^{4}}\delta \left( x-%
\frac{p^{+}}{P^{+}}\right) \alltrace\left[ iS\left( p\right)
\gamma ^{+}\frac{1}{2}\left( 1+\tau ^{3}\right) iS\left( p+\Delta \right) %
\right]
\end{equation}
and
\begin{gather}
B=\frac{2g_{\pi qq}^{2}g}{1-2g\Pi_{s}\left( P-P^{\prime }\right) }\int
\frac{d^{4}k}{\left( 2\pi \right) ^{4}}\alltrace\left[ \gamma
_{5}\tau ^{-}S\left( k-P\right) \gamma _{5}\tau ^{+}S\left( k+\Delta \right)
S\left( k\right)
+ \gamma _{5}\tau ^{+}S\left( k+P^{\prime }\right) \gamma
_{5}\tau ^{-}S\left( k+\Delta \right) S\left( k\right) \right],
\label{2.1}
\end{gather}
where $\Pi _{s}$ is the scalar vacuum polarization:
\begin{equation}
\Pi _{s}\left( P\right) =-i\int \frac{d^{4}p}{\left( 2\pi \right) ^{4}}%
\alltrace\left[ iS\left( p\right) iS\left( p-P\right)
\right] =4iN_{c}N_{f}\left[ I_{1}+\frac{1}{2}\left(
4m^{2}-P^{2}\right) I_{2}\left( P\right) \right].  \label{2.2}
\end{equation}
After some algebra we obtain using Eq. (\ref{masapi})

\begin{equation}
\mathcal{H}^{b}\left( x,\zeta ,t\right) =\frac{1}{4\pi
^{2}}N_{c}g_{\pi
qq}^{2}\theta \left( \zeta -x\right) \theta \left( x\right) \left( 1-\frac{2x%
}{\zeta }\right) C\left( t\right) \sum_{j=1}^{2}c_{j}\left[ -\ln \frac{%
m_{j}^{2}}{m^{2}}-\ln \frac{m_{j}^{2}-t\bar{y}}{m_{j}^{2}}\right]
\end{equation}%
with $\bar{y}=x(\zeta -x)/\zeta ^{2}$ and
\begin{equation}
C\left( t\right) =m^{2}\frac{\left[ \left( t-2M^{2}\right)
\,I_{3}\left(
P,P^{\prime }\right) +2I_{2}\left( P-P^{\prime }\right) \right] }{%
M^{2}\,I_{2}\left( M^{2}\right) +\left( 4m^{2}-t\right)
I_{2}\left( P-P^{\prime }\right) },
\end{equation}
which are Eqs. (\ref{njlxltzspdb}) and (\ref{C(t)}).

\section{Polynomiality.}
\label{polynomiality}

One stringent test of models is the polynomiality condition. This study
appears simpler and more physical in terms of the more symmetric
variables~\cite{Ji:1998pc}:
\begin{equation}
X=\frac{x-\zeta/2}{1-\zeta/2},\qquad\xi=\frac{\zeta}{2-\zeta}.
\label{Cambio}
\end{equation}
With this definition $X$ ranges from $-1$ to $+1$, and not from $\zeta-1$ to
$+1$, which is the range associated with $x.$ In terms of these new
variables the polynomiality condition is stated as~\cite{Ji:1998pc}:
\begin{equation}
H_{n}\left( \xi,t\right) =\int_{-1}^{+1}dX\,X^{n-1}\,\mathcal{H}\left(
X,\xi,t\right) =\sum_{i=0}^{\left[ \frac{n}{2}\right] }A_{n,2i}\left(
t\right) \,\xi^{2i}
\label{Poli}
\end{equation}
where $\left[ ...\right] $ means the integer part. Time reversal
invariance and hermiticity imply that the distributions
${H}_{n}\left( \xi,t\right) $ are even functions of
$\xi$~\cite{Ji:1998pc}.

For SED in the case $m_\pi=0$ and $t=0$, as function of the new variables,
we have
\begin{equation}
\mathcal{H}\left( X,\xi,0\right) =\left\{
\begin{array}{lcc}
\frac{6}{1+\xi}\frac{X\left( X+\xi\right) }{2\xi}, &  & -\xi< X< \xi \\
\frac{6}{1+\xi}\frac{X\left( 1-X\right) }{1-\xi}, &  & \,\xi < X< 1%
\end{array}
\right.  \label{GPDscalar}
\end{equation}
Using Eq.~(\ref{Poli}) we obtain for the coefficients,
\begin{equation}
A_{n,2i}^{SED}\left( 0\right) =\frac{6}{\left( n+1\right) \left( n+2\right) }%
.  \label{AS}
\end{equation}
This simple closed form only appears for vanishing $t$ and $m_\pi$. For non
zero values we have to proceed numerically. In table~\ref{table} we show
values of $A_{n,2i}^{SED}\left( t\right) $ for the lowest values of $n$, and
non vanishing $t$ and $m_\pi$. These results show that $A_{n,2i}^{SED}\left(
t\right) $ are not anymore independent of $i.$ We also obtain that the
effect of a physical pion mass is small, i.e. at the level of a few percent.

The definition of the coefficients in Eq.~(\ref{AS}) leads to the following
relation for their sum, which is connected with the value of the
moments~(\ref{Poli}) at $\xi=1,$
\begin{equation}
H_{2m-1}\left( 1,0\right) =H_{2m}\left( 1,0\right) ,
\end{equation}
with $m = 1,2,...$ We have verified numerically (with a precision of $%
10^{-13}) $ that this relation is also exactly satisfied for $t\neq 0$ and $%
m_{\pi }=0$:
\begin{equation}
H_{2m-1}\left( 1,t\right) =\sum_{i=0}^{m-1}A_{2m-1,2i}^{SED}\left( t\right)
=H_{2m}\left( 1,t\right) =\sum_{i=0}^{m}A_{2m,2i}^{SED}\left( t\right) ,
\label{SUMA1}
\end{equation}
For $m_{\pi }\neq 0$ the maximum value of $\xi $ never reaches 1 and
therefore we cannot connect the sum of the $A_{n,2i}$ coefficients with the
value of the moment at any physical value of $\xi $. Our numerical study
shows however that, in the case of SED, we have an approximate sum rule for
the sum of coefficients $A_{n,2i}$:
\begin{equation}
\sum_{i=0}^{m-1}A_{2m-1,2i}^{SED}\left( t\right)
-\sum_{i=0}^{m}A_{2m,2i}^{SED}\left( t\right) \simeq \mathcal{O}\left(
m_{\pi }^{2}\right)  \label{suma1a}
\end{equation}
The deviation from the exact sum rule is at the level of one percent for
$t=-10^{-5}$ GeV$^{2}$ ($-t$ small) and at the level of $10^{-3}$ for $t=-10$
GeV$^{2}$ ($-t$ large).

We now turn to the NJL model. As in the SED case, we start from
the parton distribution for $t=0$ and $m_{\pi }=0$:
\begin{equation}
\mathcal{H}\left( X,\xi ,0\right) =\frac{1}{2}\theta \left( \xi -X\right)
\,\theta \left( \xi +X\right) +\theta \left( X-\xi \right) \,\theta \left(
1-X\right)  \label{GPDnjl}
\end{equation}
Inserting Eq~(\ref{GPDnjl}) in Eq.~(\ref{Poli}) and integrating
over $X$ we obtain the coefficients,
\begin{eqnarray}
A_{n,0}^{NJL}\left( 0\right) &=&\frac{1}{n},  \notag \\
A_{n,2i}^{NJL}\left( 0\right) &=&0,\qquad 1\leq 2i\leq n-1,  \label{ANJL} \\
A_{n,n}^{NJL}\left( 0\right) &=&\frac{-1}{n}\;\;\delta _{\lbrack \frac{n}{2}%
],\frac{n}{2}}.  \notag
\end{eqnarray}
In this case, the closed forms appear only for vanishing
values of $m_{\pi }$ and $t$. Results for $t\neq 0$
and $m_{\pi }\neq 0$ have been obtained numerically and are shown
in table~\ref{table}. From these results we realize that
$A_{n,2i}^{NJL}\left( t\right) $ are not zero for $1\leq 2i\leq
n-1$. We also observe that the effect of a physical pion mass is
small.

We have found also several simple relations like
\begin{equation}
H_{2m}\left( 1,0\right) =0, \qquad\qquad
H_{2m+1}\left( 1,0\right) =H_{2m+1}\left( 0,0\right),
\end{equation}
with $m = 1,2...$. But in this case these relations are only
approximately satisfied for $t\neq 0$ and $m_{\pi }=0$ or for
$m_{\pi }\neq 0$.

We have studied numerically the contributions to $H_{n}\left( \xi,t\right)$
coming from each diagram separately. We observe that
there are independent sum rules for each diagram but they
disappear when the sum is performed. The contribution coming from
the diagram in Fig.~\ref{fighandbag} verifies the relation
\begin{equation}
\sum_{i=0}^{m}A_{2m,2i}^{NJL}\left( t\right)
-\sum_{i=0}^{m}A_{2m+1,2i}^{NJL}\left( t\right) =\mathcal{O}\left( m_{\pi
}^{2}\right)
\end{equation}
exactly for $m = 0$ and any $t,$ and in an approximate way (deviations
less than $10^{-3})$ for
$m_{\pi }=140\mathrm{MeV}$. On the other
hand, the diagram in Fig.~\ref{figsigmadiagram} gives
non-vanishing contributions only to the coefficients
$A_{n,n}^{NJL}$ for any value of $m_{\pi }$ and any value of $t.$

Regarding the numerical values of these coefficients we observe
that in the scalar model and for large $n$, $H_{n}\left(
0,0\right) $ decreases with $n$ as $1/n^{2}$ whereas $H_{n}\left(
1,0\right) $ decreases with $n$ as $1/n$, showing that
$\mathcal{H}\left( X,1,0\right) $ is more concentrated around
$X=1$ than $\mathcal{H}\left( X ,0,0\right)$.
In the NJL case for large $n$, $H_{n}\left( 0,0\right) $ and
$H_{n}\left( 1,0\right) $ decrease as $1/n.$ This shows that
$\mathcal{H}\left( X,\xi ,0\right) $  is more concentrated around
the $X=1$ region than in the scalar case. These asymptotic
behaviors appear for any value of $t$  and $m_{\pi }$.

The numerical analysis also shows that for small $n$, as $-t$
increases, the coefficients of the SED and NJL models are of the
same order, loosing the characteristic behavior of equations~(\ref{AS})
and~(\ref{ANJL}). Nevertheless we observe a regularity
in the signs: all the coefficients are positive in the scalar
model, whereas in the NJL model the coefficients $A_{n,0}$ are
positive but the rest are negative, except some small coefficients
present at small $-t$.

Finally we recall that $A_{1,0}\left( 0\right) =1$ corresponds to the charge
sum rule and $A_{2,0}\left( 0\right) =1/2$ is related to the momentum sum
rule.

\begin{table}[tbp]
\hspace*{-3cm}%
\begin{tabular}{|c|c|c|c|c|}
\hline
$n$ & 1 & 2 & 3 & 4 \\ \hline
& $A_{1,0}\left( t\right) $ & $%
\begin{array}{rr}
A_{2,0}\left( t\right)  & A_{2,2}\left( t\right)
\end{array}%
$ & $%
\begin{array}{rr}
A_{3,0}\left( t\right)  & A_{3,2}\left( t\right)
\end{array}%
$ & $%
\begin{array}{rrr}
A_{4,0}\left( t\right)  & A_{4,2}\left( t\right)  & A_{4,4}\left( t\right)
\end{array}%
$ \\ \hline
$%
\begin{array}{c}
\text{SED} \\
m_{\pi }=0\,\,\,\,\,%
\end{array}%
\,\,\,\,\,%
\begin{array}{l}
t=0 \\
t=-1 \\
t=-10%
\end{array}%
$ & $%
\begin{array}{l}
1. \\
0.629 \\
\multicolumn{1}{r}{0.240}%
\end{array}%
$ & $%
\begin{array}{ll}
0.5 & 0.5 \\
0.355 & 0.274 \\
\multicolumn{1}{r}{0.157} & \multicolumn{1}{r}{0.083}%
\end{array}%
$ & $%
\begin{array}{ll}
0.3 & 0.3 \\
0.230 & 0.187 \\
\multicolumn{1}{r}{0.113} & \multicolumn{1}{r}{0.065}%
\end{array}%
$ & $%
\begin{array}{lll}
0.2 & 0.2 & 0.2 \\
0.162 & 0.137 & 0.119 \\
\multicolumn{1}{r}{0.087} & \multicolumn{1}{r}{0.053} & \multicolumn{1}{r}{
0.039}%
\end{array}%
$ \\ \hline
$%
\begin{array}{c}
\text{SED} \\
m_{\pi }=140%
\end{array}%
$\thinspace $\,\,\,\,%
\begin{array}{l}
t=0[-10^{-5}] \\
t=-1 \\
t=-10%
\end{array}%
$ & $%
\begin{array}{l}
1. \\
0.615 \\
\multicolumn{1}{r}{0.229}%
\end{array}%
$ & $%
\begin{array}{ll}
0.5 & [0.488] \\
0.348 & 0.264 \\
0.150 & 0.079%
\end{array}%
$ & $%
\begin{array}{ll}
0.299 & [0.299] \\
0.226 & 0.183 \\
0.109 & 0.063%
\end{array}%
$ & $%
\begin{array}{lll}
0.199 & [0.201] & [0.166] \\
0.158 & 0.135 & 0.114 \\
0.083 & 0.051 & 0.037%
\end{array}%
$ \\ \hline
$%
\begin{array}{c}
\text{NJL} \\
m_{\pi }=0\,\,\,\,\,%
\end{array}%
\,\,\,\,\,\,%
\begin{array}{l}
t=0 \\
t=-1 \\
t=-10%
\end{array}%
$ & $%
\begin{array}{l}
1. \\
0.487 \\
\multicolumn{1}{r}{0.091}%
\end{array}%
$ & $%
\begin{array}{ll}
0.5 & -0.5 \\
0.336 & -0.240 \\
\multicolumn{1}{r}{0.119} & \multicolumn{1}{r}{-0.093}%
\end{array}%
$ & $%
\begin{array}{ll}
0.333 & 0. \\
0.257 & -0.049 \\
\multicolumn{1}{r}{0.118} & \multicolumn{1}{r}{-0.058}%
\end{array}%
$ & $%
\begin{array}{lll}
0.25 & 0. & -0.25 \\
0.208 & -0.030 & -0.113 \\
\multicolumn{1}{r}{0.111} & \multicolumn{1}{r}{-0.046} & \multicolumn{1}{r}{
-0.044}%
\end{array}%
$ \\ \hline
$%
\begin{array}{c}
\text{NJL} \\
m_{\pi }=140%
\end{array}%
\,\,\,\,\,%
\begin{array}{l}
t=0[-10^{-5}] \\
t=-1 \\
t=-10%
\end{array}%
$ & $%
\begin{array}{l}
1. \\
\multicolumn{1}{r}{0.482} \\
\multicolumn{1}{r}{0.090}%
\end{array}%
$ & $%
\begin{array}{ll}
0.5 & [-0.473] \\
\multicolumn{1}{r}{0.332} & \multicolumn{1}{r}{-0.230} \\
\multicolumn{1}{r}{0.116} & \multicolumn{1}{r}{-0.088}%
\end{array}%
$ & $%
\begin{array}{ll}
0.332 & [0.005] \\
\multicolumn{1}{r}{0.253} & \multicolumn{1}{r}{-0.045} \\
\multicolumn{1}{r}{0.114} & \multicolumn{1}{r}{-0.055}%
\end{array}%
$ & $%
\begin{array}{lll}
0.247 & [0.005] & [-0.236] \\
0.203 & -0.026 & -0.109 \\
\multicolumn{1}{r}{0.108} & \multicolumn{1}{r}{-0.043} & \multicolumn{1}{r}{
-0.042}%
\end{array}%
$ \\ \hline
\end{tabular}
\caption{Coefficients of the polynomial expansion. The pion mass is expressed in MeV
and $t$ is expressed in GeV$^2$.
Values between brackets correspond to $t=-10^{-5}$ GeV$^2$.
\label{table}}
\end{table}%

%


\begin{thebibliography}{99}

\bibitem[M{\"{u}}ller et~al.(1994)M{\"{u}}ller, Robaschik, Geyer, Dittes,
and Horejsi]{Muller:1994fv} \bibinfo{author}{\bibfnamefont{D.}~%
\bibnamefont{M{\"u}ller}}, \bibinfo{author}{\bibfnamefont{D.}~%
\bibnamefont{Robaschik}}, \bibinfo{author}{\bibfnamefont{B.}~%
\bibnamefont{Geyer}},
\bibinfo{author}{\bibfnamefont{F.~M.}
\bibnamefont{Dittes}}, and \bibinfo{author}{\bibfnamefont{J.}~%
\bibnamefont{Horejsi}}, \bibinfo{journal}{Fortschr. Phys.} \textbf{%
\bibinfo{volume}{42}}, \bibinfo{pages}{101} (\bibinfo{year}{1994}).

\bibitem[Ji(1997)]{Ji:1997nm}
\bibinfo{author}{\bibfnamefont{X.-D.}
\bibnamefont{Ji}}, \bibinfo{journal}{Phys. Rev.} \textbf{%
\bibinfo{volume}{D55}}, \bibinfo{pages}{7114} (\bibinfo{year}{1997}).

\bibitem[Radyushkin(1997)]{Radyushkin:1997ki} \bibinfo{author}{%
\bibfnamefont{A.~V.} \bibnamefont{Radyushkin}}, \bibinfo{journal}{Phys. Rev.}
\textbf{\bibinfo{volume}{D56}}, \bibinfo{pages}{5524} (\bibinfo{year}{1997}%
). 

\bibitem[Ji(1998)]{Ji:1998pc}
\bibinfo{author}{\bibfnamefont{X.-D.}
\bibnamefont{Ji}}, \bibinfo{journal}{J.
  Phys.} \textbf{\bibinfo{volume}{G24}}, \bibinfo{pages}{1181} (%
\bibinfo{year}{1998}). 

\bibitem[Diehl et~al.(1999)Diehl, Feldmann, Jakob, and Kroll]{Diehl:1998kh} %
\bibinfo{author}{\bibfnamefont{M.}~\bibnamefont{Diehl}}, \bibinfo{author}{%
\bibfnamefont{T.}~\bibnamefont{Feldmann}}, \bibinfo{author}{%
\bibfnamefont{R.}~\bibnamefont{Jakob}}, and \bibinfo{author}{%
\bibfnamefont{P.}~\bibnamefont{Kroll}}, \bibinfo{journal}{Eur. Phys. J.}
\textbf{\bibinfo{volume}{C8}}, \bibinfo{pages}{409} (\bibinfo{year}{1999}).

\bibitem[Diehl et~al.(2001)Diehl, Feldmann, Jakob, and Kroll]{Diehl:2000xz} %
\bibinfo{author}{\bibfnamefont{M.}~\bibnamefont{Diehl}}, \bibinfo{author}{%
\bibfnamefont{T.}~\bibnamefont{Feldmann}}, \bibinfo{author}{%
\bibfnamefont{R.}~\bibnamefont{Jakob}}, and \bibinfo{author}{%
\bibfnamefont{P.}~\bibnamefont{Kroll}}, \bibinfo{journal}{Nucl. Phys.}
\textbf{\bibinfo{volume}{B596}}, \bibinfo{pages}{33} (\bibinfo{year}{2001}).

\bibitem[Polyakov and Weiss(1999)]{Polyakov:1999gs} \bibinfo{author}{%
\bibfnamefont{M.~V.} \bibnamefont{Polyakov}} and \bibinfo{author}{%
\bibfnamefont{C.}~\bibnamefont{Weiss}}, \bibinfo{journal}{Phys. Rev.}
\textbf{\bibinfo{volume}{D60}}, \bibinfo{pages}{114017} (\bibinfo{year}{1999}%
). 

\bibitem[Diehl et~al.(2000)Diehl, Feldmann, Kroll, and Vogt]{Diehl:1999ek} %
\bibinfo{author}{\bibfnamefont{M.}~\bibnamefont{Diehl}}, \bibinfo{author}{%
\bibfnamefont{T.}~\bibnamefont{Feldmann}}, \bibinfo{author}{%
\bibfnamefont{P.}~\bibnamefont{Kroll}}, and \bibinfo{author}{%
\bibfnamefont{C.}~\bibnamefont{Vogt}}, \bibinfo{journal}{Phys. Rev.} \textbf{%
\bibinfo{volume}{D61}}, \bibinfo{pages}{074029} (\bibinfo{year}{2000}).

\bibitem[Ji et~al.(1997)Ji, Melnitchouk, and Song]{Ji:1997gm} %
\bibinfo{author}{\bibfnamefont{X.-D.} \bibnamefont{Ji}}, \bibinfo{author}{%
\bibfnamefont{W.}~\bibnamefont{Melnitchouk}}, and \bibinfo{author}{%
\bibfnamefont{X.}~\bibnamefont{Song}}, \bibinfo{journal}{Phys. Rev.} \textbf{%
\bibinfo{volume}{D56}}, \bibinfo{pages}{5511} (\bibinfo{year}{1997}).

\bibitem[Anikin et~al.(2002)Anikin, Binosi, Medrano, Noguera, and Vento]%
{Anikin:2001zv} \bibinfo{author}{\bibfnamefont{I.~V.} \bibnamefont{Anikin}}, %
\bibinfo{author}{\bibfnamefont{D.}~\bibnamefont{Binosi}}, %
\bibinfo{author}{\bibfnamefont{R.}~\bibnamefont{Medrano}}, %
\bibinfo{author}{\bibfnamefont{S.}~\bibnamefont{Noguera}}, and %
\bibinfo{author}{\bibfnamefont{V.}~\bibnamefont{Vento}}, %
\bibinfo{journal}{Eur. Phys. J.} \textbf{\bibinfo{volume}{A14}}, %
\bibinfo{pages}{95} (\bibinfo{year}{2002}).

\bibitem[Petrov et~al.(1998)]{Petrov:1998kf} \bibinfo{author}{%
\bibfnamefont{V.~Y.} \bibnamefont{Petrov}} et~al.,
\bibinfo{journal}{Phys.
Rev.} \textbf{\bibinfo{volume}{D57}}, \bibinfo{pages}{4325} (%
\bibinfo{year}{1998}). 

\bibitem[Penttinen et~al.(2000)Penttinen, Polyakov, and Goeke]%
{Penttinen:1999th} \bibinfo{author}{\bibfnamefont{M.}~%
\bibnamefont{Penttinen}},
\bibinfo{author}{\bibfnamefont{M.~V.}
\bibnamefont{Polyakov}}, and \bibinfo{author}{\bibfnamefont{K.}~%
\bibnamefont{Goeke}}, \bibinfo{journal}{Phys. Rev.} \textbf{%
\bibinfo{volume}{D62}}, \bibinfo{pages}{014024} (\bibinfo{year}{2000}).

\bibitem[Scopetta and Vento(2002)]{Scopetta:2002mz} \bibinfo{author}{%
\bibfnamefont{S.}~\bibnamefont{Scopetta}} and \bibinfo{author}{%
\bibfnamefont{V.}~\bibnamefont{Vento}}, \bibinfo{journal}{Nucl. Phys.}
\textbf{\bibinfo{volume}{A711}}, \bibinfo{pages}{190} (\bibinfo{year}{2002}%
). 

\bibitem[Scopetta and Vento(2003)]{Scopetta:2002xq} \bibinfo{author}{%
\bibfnamefont{S.}~\bibnamefont{Scopetta}} and \bibinfo{author}{%
\bibfnamefont{V.}~\bibnamefont{Vento}}, \bibinfo{journal}{Eur. Phys. J.}
\textbf{\bibinfo{volume}{A16}}, \bibinfo{pages}{527} (\bibinfo{year}{2003}).

\bibitem[Klevansky(1992)]{Klevansky:1992qe} \bibinfo{author}{%
\bibfnamefont{S.~P.} \bibnamefont{Klevansky}},
\bibinfo{journal}{Rev. Mod.
Phys.} \textbf{\bibinfo{volume}{64}}, \bibinfo{pages}{649} (%
\bibinfo{year}{1992}).

\bibitem[Bijnens(1996)]{Bijnens:1996ww} \bibinfo{author}{\bibfnamefont{J.}~%
\bibnamefont{Bijnens}}, \bibinfo{journal}{Phys. Rept.} \textbf{%
\bibinfo{volume}{265}}, \bibinfo{pages}{369} (\bibinfo{year}{1996}).

\bibitem[Golec-Biernat and Martin(1999)]{Golec-Biernat:1998ja} %
\bibinfo{author}{\bibfnamefont{K.~J.} \bibnamefont{Golec-Biernat}} and
\bibinfo{author}{\bibfnamefont{A.~D.}
  \bibnamefont{Martin}}, \bibinfo{journal}{Phys. Rev.} \textbf{%
\bibinfo{volume}{D59}}, \bibinfo{pages}{014029} (\bibinfo{year}{1999}).

\bibitem[Guichon and Vanderhaeghen(1998)]{Guichon:1998xv} %
\bibinfo{author}{\bibfnamefont{P.~A.~M.} \bibnamefont{Guichon}} and %
\bibinfo{author}{\bibfnamefont{M.}~\bibnamefont{Vanderhaeghen}}, %
\bibinfo{journal}{Prog. Part. Nucl. Phys.} \textbf{\bibinfo{volume}{41}}, %
\bibinfo{pages}{125} (\bibinfo{year}{1998}).

\bibitem[Itzykson and Zuber(1985)]{Itzykson:1985} \bibinfo{author}{%
\bibfnamefont{C.}~\bibnamefont{Itzykson}} and \bibinfo{author}{%
\bibfnamefont{J.~B.} \bibnamefont{Zuber}}, \emph{%
\bibinfo{title}{Quantum
{F}ield {T}heory}} (\bibinfo{publisher}{McGraw-Hill}, \bibinfo{year}{1985}).

\bibitem[Desplanques et~al.(2002)Desplanques, Theu{\ss }l, and Noguera]%
{Desplanques:2001ze} \bibinfo{author}{\bibfnamefont{B.}~%
\bibnamefont{Desplanques}}, \bibinfo{author}{\bibfnamefont{L.}~%
\bibnamefont{Theu{\ss}l}}, and \bibinfo{author}{\bibfnamefont{S.}~%
\bibnamefont{Noguera}}, \bibinfo{journal}{Phys. Rev.} \textbf{%
\bibinfo{volume}{C65}}, \bibinfo{pages}{038202} (\bibinfo{year}{2002}).

\bibitem[Amghar et~al.(2003)Amghar, Desplanques, and Theu{\ss }l]%
{Amghar:2002jx} \bibinfo{author}{\bibfnamefont{A.}~\bibnamefont{Amghar}}, %
\bibinfo{author}{\bibfnamefont{B.}~\bibnamefont{Desplanques}}, and %
\bibinfo{author}{\bibfnamefont{L.}~\bibnamefont{Theu{\ss}l}}, %
\bibinfo{journal}{Nucl. Phys.} \textbf{\bibinfo{volume}{A714}}, %
\bibinfo{pages}{213} (\bibinfo{year}{2003}). 

\bibitem[Tiburzi and Miller(2003)]{Tiburzi:2002tq} \bibinfo{author}{%
\bibfnamefont{B.~C.} \bibnamefont{Tiburzi}} and \bibinfo{author}{%
\bibfnamefont{G.~A.} \bibnamefont{Miller}}, \bibinfo{journal}{Phys. Rev.}
\textbf{\bibinfo{volume}{D67}}, \bibinfo{pages}{113004} (\bibinfo{year}{2003}%
). 

\bibitem[Tiburzi and Miller(2001)]{Tiburzi:2001ta} \bibinfo{author}{%
\bibfnamefont{B.~C.} \bibnamefont{Tiburzi}} and \bibinfo{author}{%
\bibfnamefont{G.~A.} \bibnamefont{Miller}}, \bibinfo{journal}{Phys. Rev.}
\textbf{\bibinfo{volume}{C64}}, \bibinfo{pages}{065204} (\bibinfo{year}{2001}%
). 

\bibitem[Nambu and Jona-Lasinio(1961{a})]{Nambu:1961tp} \bibinfo{author}{%
\bibfnamefont{Y.}~\bibnamefont{Nambu}} and \bibinfo{author}{%
\bibfnamefont{G.}~\bibnamefont{Jona-Lasinio}}, \bibinfo{journal}{Phys. Rev.}
\textbf{\bibinfo{volume}{122}}, \bibinfo{pages}{345} (\bibinfo{year}{1961}{%
\natexlab{a}}).

\bibitem[Nambu and Jona-Lasinio(1961{b})]{Nambu:1961fr} \bibinfo{author}{%
\bibfnamefont{Y.}~\bibnamefont{Nambu}} and \bibinfo{author}{%
\bibfnamefont{G.}~\bibnamefont{Jona-Lasinio}}, \bibinfo{journal}{Phys. Rev.}
\textbf{\bibinfo{volume}{124}}, \bibinfo{pages}{246} (\bibinfo{year}{1961}{%
\natexlab{b}}).

\bibitem[Vogl and Weise(1991)]{Vogl:1991qt} \bibinfo{author}{%
\bibfnamefont{U.}~\bibnamefont{Vogl}} and \bibinfo{author}{%
\bibfnamefont{W.}~\bibnamefont{Weise}},
\bibinfo{journal}{Prog. Part. Nucl.
Phys.} \textbf{\bibinfo{volume}{27}}, \bibinfo{pages}{195} (%
\bibinfo{year}{1991}).

\bibitem[Anikin et~al.(2000{a})Anikin, Dorokhov, Maksimov, Tomio, and Vento]%
{Anikin:2000th} \bibinfo{author}{\bibfnamefont{I.~V.} \bibnamefont{Anikin}}, %
\bibinfo{author}{\bibfnamefont{A.~E.} \bibnamefont{Dorokhov}}, %
\bibinfo{author}{\bibfnamefont{A.~E.} \bibnamefont{Maksimov}}, %
\bibinfo{author}{\bibfnamefont{L.}~\bibnamefont{Tomio}}, and %
\bibinfo{author}{\bibfnamefont{V.}~\bibnamefont{Vento}}, %
\bibinfo{journal}{Nucl. Phys.} \textbf{\bibinfo{volume}{A678}}, %
\bibinfo{pages}{175} (\bibinfo{year}{2000}{\natexlab{a}}).

\bibitem[Vogt(2001)]{Vogt:2001if} \bibinfo{author}{\bibfnamefont{C.}~%
\bibnamefont{Vogt}}, \bibinfo{journal}{Phys.
  Rev.} \textbf{\bibinfo{volume}{D64}}, \bibinfo{pages}{057501} (%
\bibinfo{year}{2001}). 

\bibitem[Choi et~al.(2001)Choi, Ji, and Kisslinger]{Choi:2001fc} %
\bibinfo{author}{\bibfnamefont{H.-M.} \bibnamefont{Choi}}, %
\bibinfo{author}{\bibfnamefont{C.-R.} \bibnamefont{Ji}}, and %
\bibinfo{author}{\bibfnamefont{L.~S.} \bibnamefont{Kisslinger}}, %
\bibinfo{journal}{Phys. Rev.} \textbf{\bibinfo{volume}{D64}}, %
\bibinfo{pages}{093006} (\bibinfo{year}{2001}).

\bibitem[Kisslinger et~al.(2001)Kisslinger, Choi, and Ji]{Kisslinger:2001gw} %
\bibinfo{author}{\bibfnamefont{L.~S.} \bibnamefont{Kisslinger}}, %
\bibinfo{author}{\bibfnamefont{H.-M.} \bibnamefont{Choi}}, and %
\bibinfo{author}{\bibfnamefont{C.-R.} \bibnamefont{Ji}}, %
\bibinfo{journal}{Phys. Rev.} \textbf{\bibinfo{volume}{D63}}, %
\bibinfo{pages}{113005} (\bibinfo{year}{2001}).

\bibitem[Davidson and Ruiz~Arriola(2002)]{Davidson:2001cc} %
\bibinfo{author}{\bibfnamefont{R.~M.} \bibnamefont{Davidson}} and %
\bibinfo{author}{\bibfnamefont{E.}~\bibnamefont{Ruiz~Arriola}}, %
\bibinfo{journal}{Acta Phys. Polon.} \textbf{\bibinfo{volume}{B33}}, %
\bibinfo{pages}{1791} (\bibinfo{year}{2002}). 

\bibitem[Davidson and Ruiz~Arriola(1995)]{Davidson:1995uv} %
\bibinfo{author}{\bibfnamefont{R.~M.} \bibnamefont{Davidson}} and %
\bibinfo{author}{\bibfnamefont{E.}~\bibnamefont{Ruiz~Arriola}}, %
\bibinfo{journal}{Phys. Lett.} \textbf{\bibinfo{volume}{B348}}, %
\bibinfo{pages}{163} (\bibinfo{year}{1995}).

\bibitem[Weigel et~al.(1999)Weigel, Ruiz~Arriola, and Gamberg]%
{Weigel:1999pc} \bibinfo{author}{\bibfnamefont{H.}~\bibnamefont{Weigel}}, %
\bibinfo{author}{\bibfnamefont{E.}~\bibnamefont{Ruiz~Arriola}}, and
\bibinfo{author}{\bibfnamefont{L.~P.}
  \bibnamefont{Gamberg}}, \bibinfo{journal}{Nucl. Phys.} \textbf{%
\bibinfo{volume}{B560}}, \bibinfo{pages}{383} (\bibinfo{year}{1999}).

\bibitem[Shigetani et~al.(1993)Shigetani, Suzuki, and Toki]%
{Shigetani:1993dx} \bibinfo{author}{\bibfnamefont{T.}~%
\bibnamefont{Shigetani}}, \bibinfo{author}{\bibfnamefont{K.}~%
\bibnamefont{Suzuki}}, and \bibinfo{author}{\bibfnamefont{H.}~%
\bibnamefont{Toki}}, \bibinfo{journal}{Phys. Lett.} \textbf{%
\bibinfo{volume}{B308}}, \bibinfo{pages}{383} (\bibinfo{year}{1993}).

\bibitem[Polyakov(1999)]{Polyakov:1998ze} \bibinfo{author}{%
\bibfnamefont{M.~V.} \bibnamefont{Polyakov}}, \bibinfo{journal}{Nucl. Phys.}
\textbf{\bibinfo{volume}{B555}}, \bibinfo{pages}{231} (\bibinfo{year}{1999}%
). 

\bibitem[Bernard and Meissner(1988)]{Bernard:1988bx} \bibinfo{author}{%
\bibfnamefont{V.}~\bibnamefont{Bernard}} and \bibinfo{author}{%
\bibfnamefont{U.~G.} \bibnamefont{Meissner}},
\bibinfo{journal}{Phys. Rev.
Lett.} \textbf{\bibinfo{volume}{61}}, \bibinfo{pages}{2296} (%
\bibinfo{year}{1988}).

\bibitem[Schulze(1994)]{Schulze:1994fy} \bibinfo{author}{%
\bibfnamefont{H.~J.} \bibnamefont{Schulze}}, \bibinfo{journal}{J. Phys.}
\textbf{\bibinfo{volume}{G20}}, \bibinfo{pages}{531} (\bibinfo{year}{1994}).

\bibitem[Tiburzi and Miller(2002)]{Tiburzi:2001je} \bibinfo{author}{%
\bibfnamefont{B.~C.} \bibnamefont{Tiburzi}} and \bibinfo{author}{%
\bibfnamefont{G.~A.} \bibnamefont{Miller}}, \bibinfo{journal}{Phys. Rev.}
\textbf{\bibinfo{volume}{D65}}, \bibinfo{pages}{074009} (\bibinfo{year}{2002}%
). 

\bibitem[Praszalowicz and Rostworowski(2002)]{Praszalowicz:2002ct} %
\bibinfo{author}{\bibfnamefont{M.}~\bibnamefont{Praszalowicz}} and %
\bibinfo{author}{\bibfnamefont{A.}~\bibnamefont{Rostworowski}} (%
\bibinfo{year}{2002}), \eprint[http://arXiv.org/abs]{hep-ph/0205177}.

\bibitem[Dorokhov and Tomio(2000)]{Dorokhov:2000gu} \bibinfo{author}{%
\bibfnamefont{A.~E.} \bibnamefont{Dorokhov}} and \bibinfo{author}{%
\bibfnamefont{L.}~\bibnamefont{Tomio}}, \bibinfo{journal}{Phys. Rev.}
\textbf{\bibinfo{volume}{D62}}, \bibinfo{pages}{014016} (\bibinfo{year}{2000}%
).

\bibitem[Anikin et~al.(2000{b})Anikin, Dorokhov, and Tomio]{Anikin:1999cx} %
\bibinfo{author}{\bibfnamefont{I.~V.} \bibnamefont{Anikin}}, %
\bibinfo{author}{\bibfnamefont{A.~E.} \bibnamefont{Dorokhov}}, and %
\bibinfo{author}{\bibfnamefont{L.}~\bibnamefont{Tomio}}, %
\bibinfo{journal}{Phys. Lett.} \textbf{\bibinfo{volume}{B475}}, %
\bibinfo{pages}{361} (\bibinfo{year}{2000}{\natexlab{b}}).

\bibitem[Ruiz~Arriola and Broniowski(2002)]{RuizArriola:2002bp} %
\bibinfo{author}{\bibfnamefont{E.}~\bibnamefont{Ruiz~Arriola}} and %
\bibinfo{author}{\bibfnamefont{W.}~\bibnamefont{Broniowski}}, %
\bibinfo{journal}{Phys. Rev.} \textbf{\bibinfo{volume}{D66}}, %
\bibinfo{pages}{094016} (\bibinfo{year}{2002}). 

\bibitem['t~Hooft(1974)]{'tHooft:1974hx} \bibinfo{author}{\bibfnamefont{G.}~%
\bibnamefont{'t~Hooft}}, \bibinfo{journal}{Nucl. Phys.} \textbf{%
\bibinfo{volume}{B75}}, \bibinfo{pages}{461} (\bibinfo{year}{1974}).

\bibitem[M{\"{u}}ller(1995)]{Muller:1995cn} \bibinfo{author}{%
\bibfnamefont{D.}~\bibnamefont{M{\"u}ller}}, \bibinfo{journal}{Phys. Rev.}
\textbf{\bibinfo{volume}{D51}}, \bibinfo{pages}{3855} (\bibinfo{year}{1995}%
). 
\end{thebibliography}
\ifx\csname natexlab\endcsname\relax \fi\expandafter\ifx\csname bibnamefont%
\endcsname\relax \fi\expandafter\ifx\csname bibfnamefont\endcsname\relax \fi%
\expandafter\ifx\csname citenamefont\endcsname\relax \fi\expandafter\ifx%
\csname url\endcsname\relax \fi\expandafter\ifx\csname urlprefix\endcsname%
\relax \fi\providecommand{\bibinfo}[2]{#2} \providecommand{\eprint}[2][]{%
\url{#2}}

\end{document}